\documentclass[a4paper,11pt]{article}
\pdfoutput=1
\usepackage{jcappub}
\usepackage{epsf}
\usepackage[utf8]{inputenc}
\usepackage{amstext}
\usepackage{graphicx}
\usepackage{amssymb}
\usepackage{amsmath}
\usepackage{multirow}
\usepackage{amsthm}

\title{\boldmath Quantifying the evidence for the current speed-up of the Universe with low and intermediate-redshift data. A more model-independent approach}

\author{Adri\`a G\'omez-Valent}

\affiliation{Departament de F\'isica Qu\`antica i Astrof\'isica, and Institut
de Ciències del Cosmos, Univ.\ de Barcelona, Av.\ Diagonal 647, E-08028 Barcelona,
Catalonia}

\emailAdd{adriagova@fqa.ub.edu}

\abstract{%
We reconstruct in this paper the deceleration and jerk parameters as functions of the cosmological redshift from data on cosmic chronometers (CCH), baryon acoustic oscillations (BAOs), and the Pantheon+MCT compilation of supernovae of Type Ia (SnIa). The reconstruction is carried out with the Weighted Function Regression method, previously introduced by G\'omez-Valent \& Amendola (2018). It improves the usual cosmographic approach by automatically implementing Occam's razor criterion. This makes our procedure to be more free of model and parametrization dependencies than many other analyses in the literature. The reconstructed functions are fully compatible with the predictions for the concordance model. In addition, we also discuss the confidence level at which we can claim that the Universe (assumed to be flat, homogeneous and isotropic) is currently accelerating. According to Jeffreys' scale and jargon, we find moderate evidence in favor of such speed-up using the data on SnIa+CCH, and very strong one when we also use data on BAOs. The measured current value of the deceleration parameter in the latter case reads $q_0\sim -0.60\pm 0.10$, and for the deceleration-acceleration transition redshift we find $z_t\sim 0.80\pm 0.10$. The former is $\sim 6\sigma$ away from $0$. This is in stark contrast, for instance, with the $\sim 17\sigma$ that are found in the context of the flat $\Lambda$CDM even without including the BAOs data. This indicates that cosmography and Occam's razor criterion play a crucial role in this discussion, and that estimating the evidence for positive acceleration only in the framework of a particular cosmological model or parametrization is clearly insufficient.}

\arxivnumber{arXiv:1810.02278}

\begin{document}
\maketitle
\flushbottom

\section{Introduction}
Is the Universe currently undergoing a positive acceleration phase? Certainly, we can only reply this question within the margins of precision and accuracy set by the cosmological observations, i.e. the most that we can do is to answer yes or no with the confidence level permitted by the data at our disposal. Thus, it would be better to reformulate the question in an alternative way: what is the evidence in favor of the current speed-up of the Universe according to the existing cosmological data? This work is mainly focused on answering this pivotal question using low and intermediate-redshift data, trying to do it from a very skeptical perspective and removing from the analysis as much as possible the model-dependencies that could eventually bias our final answer. These model-dependencies are actually plaguing many other works in the literature, which address the problem either in the framework of concrete cosmological models, using particular parametrizations of the deceleration parameter or the jerk (cf. formulas (\ref{eq:decelerationParam}) and (\ref{eq:jerksnap1}), respectively, and Sect. \ref{sec:cosmography}), or even truncated cosmographical series describing the luminosity distance or the Hubble function in which the highest order of the expansion is fixed to a concrete value (see the list of references below). None of these analyses are model-independent in a strict sense, and therefore neither the derived confidence regions for the current value of the deceleration parameter, which describes the acceleration status of the Universe at present. It turns out that the evidence that is obtained in favor of a positive-accelerated Universe depends very strongly on the model or parametrization that is assumed in the analysis (we will show this explicitly in Sect. \ref{sec:Motivation}), so the evidence for positive acceleration that is obtained from the data can only carry an absolute and hence powerful statistical meaning when it is inferred in a full-fledged model-independent way. Otherwise, it only tells us what is the evidence in the concrete model or parametrization employed in that particular study, and thus we are not in a position to answer our question in a completely secure manner. The point is that there does not exist any completely model-independent method to reconstruct the full shape of a particular cosmographical function from observations. One is always forced to make some assumptions that are there in a more or less subtle way, even in the so-called non-parametric approaches. Thus, we must be humble enough to admit that we can only try to use reconstruction methods as model-independent as possible and try to minimize the number of assumptions. This work aims to alleviate this situation and pave the way to the obtention of a more model-independent determination of the cosmographical functions, but before entering the details of our reconstruction method let us review some of the studies on these issues and corresponding results that one can find in the literature on the subject, which is quite vast.

The first important hints of positive acceleration were reported in the late nineties by the High-Z Supernova Search Team \cite{SNIaRiess} and the Supernova Cosmology Project \cite{SNIaPerl} collaborations, from the first measurements of the apparent magnitude of supernovae of Type Ia (SnIa) at high redshifts. The samples contained individuals up to $z= 0.97$ and $0.83$, respectively, and also included low-redshift SnIa of $z\lesssim 0.15$ from the Cal\'an/Tololo Supernova Survey and, in the first case, from the CfA sample too. In the context of the non-flat $\Lambda$CDM model they found the probability for the presence of a positive cosmological constant ($\Lambda$) in Einstein's field equations to be $P(\Lambda>0)=99\%$, and restricting the analysis to the purely flat $\Lambda$CDM, i.e. setting the current spatial curvature density $\Omega_k^{(0)}=0$, they found $\sim 3\sigma$ evidence in favor of the current positive acceleration of the Universe. Subsequent studies incorporated the data of even higher-redshift SnIa discovered with the Hubble Space Telescope (HST) \cite{Riess2001,TurnerRiess2002,Knop2003,Riess2004}, at $z\gtrsim 1$. This allowed to also find compelling evidence for a deceleration-acceleration transition at $z_t\sim 0.5$. For instance, in \cite{Riess2004} the authors made use of a parametrization for the deceleration parameter of the form $q(z)=q_0+q_1z$, with $q_0=q(0)$ and $q_1=\frac{dq}{dz}$. The deceleration parameter is directly related to the second derivative of the scale factor with respect to the cosmic time $t$ \cite{Sandage1970,Weinberg1972},
\begin{equation}\label{eq:decelerationParam}
q=-\frac{\ddot{a}}{aH^2}\,,
\end{equation}
with $H=\dot{a}/a$ being the Hubble function and the dot denoting such derivative. Considering a flat Universe they obtained $P(q_0<0)=99.2\%$ and $P(q_1>0)=99.8\%$, and thus important evidence in favor of the current positive acceleration of the Cosmos and the existence of a deceleration-acceleration transition point in the past, more concretely at $z_t=0.46\pm 0.13$, with $q(z_t)=0$. 

These pioneering studies made possible the first accurate estimations of the deceleration parameter of the history. Therein the authors explored two of the routes that have been later on subsequently revisited by the cosmological community once and again to infer the evidence for a negative $q_0$ and the existence of a transition redshift, using different data sources. In e.g. \cite{SNIaRiess,SNIaPerl,Knop2003} the authors assumed concrete cosmological models, mainly the flat and non-flat $\Lambda$CDM, and derived according to the data available at the time and in that particular cosmological scenarios the confidence intervals for $q_0$. This approach makes direct use of the gravitational field equations in a particular theoretical setting, where all the sources of the energy-momentum tensor and/or the deviations from standard General Relativity (GR) are specified beforehand. Conversely, the authors of \cite{TurnerRiess2002,Riess2004} directly parametrized the deceleration parameter without focusing in any concrete cosmological model and hence tried to orient their analyses in a more ``cosmographical'' way, although of course such parametrizations are not free of model-dependencies by definition, as we will explicitly show in this paper. 

Using cosmography (also dubbed cosmokinetics or Friedmannless Cosmology) one can extract kinematic information about the Universe from measurements of cosmological distances by only assuming the Cosmological Principle (CP), which is clearly fulfilled at very large (cosmological) scales. Cosmic microwave background (CMB) observations and the inflationary paradigm also allow us to consider a flat Universe, which helps to break important degeneracies in the cosmographical framework, e.g. between the jerk and $\Omega_k^{(0)}$ \cite{Visser2004,Visser2005,DunsbyLuongo2016}. In this geometrical approach one does not need to introduce any assumption on the metric theory of gravity or the matter-energy content of the Universe, which is something very positive. Cosmography was boosted thanks to the papers by Visser \cite{Visser2004,Visser2005}, in which he extended the cosmographical formalism of previous works (see e.g. \cite{Sandage1970,Weinberg1972}) to include also higher order terms in the expansion of the cosmological distances, as the jerk and the snap (which we will discuss in detail later on). Although the cosmographical methodology cannot shed much light on the ultimate cause of the positive acceleration of the Universe, some parameters as the jerk can be employed as direct tests of the $\Lambda$CDM and the potential time-variation of the dark energy (DE) density, see e.g. \cite{Sahni2003,Blandford2005}. Modified gravity theories and violations of the CP in standard GR can also make the jerk to deviate from the $\Lambda$CDM value, see e.g. \cite{Capozziello2008,Tedesco2018}. As mentioned before, cosmography can certainly help us to answer important questions related to the kinematic properties of the Universe without relying on a particular cosmological model, but this statement should be actually duly qualified. Cosmographical expansions of cosmological distances are still parametrizations. They are truncated Taylor series developed around e.g. $z=0$. Choosing the concrete order at which the series is cut can be tricky and the derived constraints for the various parameters involved in the expansion can be highly dependent on this choice, as was already noted in \cite{ElgaroyMultamaki2006}, and also discussed in \cite{Capozziello2011}. We can of course proceed applying some model-selection criteria based on: the computation of exact Bayesian evidence, see e.g. \cite{DEbook,AmendolaNotes}; the Akaike or Bayesian criteria \cite{Akaike,Schwarz}; or even make use of the reduced chi-squared statistic. Nevertheless, regardless of how we select the ``right'' order of the cosmographical expansion according to the existing data the following questions should be still settled down: how should we proceed if two nested cosmographical expansions behaved very similarly in terms of model selection criteria. Should we choose the one with e.g. closest reduced chi-squared statistic to $1$, and throw the other one away? Would we be legitimated to do this, independently of how close the two expansions behaved in practice? Wouldn't we be loosing precious statistical information then? We firmly believe that the usual cosmographical method must be improved in order to deal with all these subtle points, and try to remove the remaining degree of subjectivity that is inherent to the choice of the highest order of the cosmographical expansion. We will do this through the so-called Weighted Function Regression (WFR) method, which was already introduced in \cite{GomezAmendola2018}\footnote{In \cite{GomezAmendola2018} we called this method Weighted {\it Polynomial} Regression instead of Weighted {\it Function} Regression, just because in that paper we used polynomials for the basis functions. In this work, though, we will also use non-polynomial expressions (see Sect. \ref{sec:cosmography}), so this change in the name is needed.} to reconstruct the Hubble function using data on $H(z_i)$ at different redshifts extracted from cosmic chronometers (CCH) with the differential-age technique, together with the data from the SnIa of the Pantheon compilation and the HST CANDELS and CLASH Multy-Cycle Treasury (MCT) programs. In this paper we will make use of the same data sources to study the two next-to-leading order terms in the expansion of the scale factor, i.e. the deceleration and jerk parameters. We will also study the impact of some intermediate-redshift data obtained from the analysis of baryon acoustic oscillations (BAOs). The WFR method implements in practice the Occam razor criterion, not by selecting only one cosmographical expansion among the various (nested) alternatives, but incorporating all of them consistently in our calculations using appropriate and rigorous Bayesian tools. This allows us to reconstruct the aforesaid cosmographical functions in a fairer and more model and parametrization-independent way. We hope to improve thereby the methodology applied and the constraints reported on e.g., $q_0$ or $z_t$, in many other works in the literature, as those based on concrete parametrizations of the deceleration or the jerk parameters \cite{TurnerRiess2002,Riess2004,ElgaroyMultamaki2006,ShapiroTurner2006,Rapetti2007,GongWang2007,Ishida2008,CunhaLima2008,Guimaraes2009,MortsellClarkson2009,Xu2009,Cunha2009,Lu2011,GuimaraesLima2011,Giostri2012,Nair2012,Zhai2013,Akarsu2014,Mukherjee2016,Vargas2016,MamonDas2017,Jesus2018,MamonBamba2018,Amirhashchi2018}, truncated cosmographical series \cite{CattoenVisser2007a,CattoenVisser2007b,Xu2009,Vitagliano2010,Luongo2011,Xu2011,RubinHayden2016,Jesus2018,Dutta2018,Heneka2018,LiDuXu2019}, alternative expansions of the luminosity distance \cite{SemizCacimbel2015}, or specific cosmological models, including also various parametrizations of the DE density or the DE equation of state (EoS) parameter, see e.g. \cite{SNIaRiess,SNIaPerl,Riess2001,TurnerRiess2002,Knop2003,Riess2004,Rapetti2007,Nielsen2016,RubinHayden2016,Ringermacher2016,Haridasu2017,Mamon2018}. Other authors have also applied alternative techniques to reconstruct the expansion history of the Universe in a model-independent way and derive constraints on the deceleration parameter, e.g. using the smoothing method of Refs. \cite{Shafieloo2006,Shafieloo2007,Shafieloo2012}, principal component analyses \cite{ShapiroTurner2006,MortsellClarkson2009}, Gaussian processes \cite{Haridasu2018}, or piecewise natural cubic splines \cite{Tutusaus2018}. These methods are interesting and useful, but also have their own drawbacks. We deem that the WFR method rises as a good alternative to put objective and fair constraints to the most relevant kinematic functions, by using low and intermediate-redshift data. 

This paper is organized as follows. In Sect. \ref{sec:Data} we describe in detail the data sets that we employ in the reconstruction of the deceleration and jerk parameters. In Sect. \ref{sec:WFR} we motivate and explain the WFR method after introducing some basic elements of cosmography that we need in order to apply this reconstruction technique. In Sect. \ref{sec:results} we present and discuss our results, including the plots with the main reconstructed cosmographical functions and some tables. We finally present our conclusions in Sect. \ref{sec:conclusions}.

%%%%%%%%%%%%%%%%%%%%%%%%%%%%%%%%%%%%%%%%%%%%%%%%%%%%%%%%%%%%%%
%%%%%%%%%%%%%%%%%%%%%%%%%%%%%%%%%%%%%%%%%%%%%%%%%%%%%%%%%%%%%%

\section{The data sets}\label{sec:Data}

In this section we list the data that we employ in our reconstruction of the various cosmographical functions we are interested in with the weighted function regression method. We also provide the corresponding references, discuss the model dependencies and assumptions behind these data, and the way they are introduced in our analysis in order to: (i) mitigate as much as possible the effect of some of these model-dependencies; (ii) incorporate unaccounted systematic uncertainties that were not taken into account in the original references; and (iii) ease the computation of evidences and the practical implementation of the WFR method. As mentioned already at the title and abstract levels, we make use of only low and intermediate-redshift data, i.e. at $z\lesssim 2.5$. The reason is double. On the one hand, the data on CCH and SnIa are cosmology-independent, and the data on $H(z_i)$ extracted from the radial component of anisotropic BAOs can be dealt with also in a way such that the model-dependencies can be strongly suppressed, as we will explain in Sect. \ref{sec:2p2}. On the other hand, the cosmographic approach works optimally only with data at this approximate redshift range. It is difficult to include in the analysis e.g. the CMB data, since the latter would force us to consider higher order terms in the cosmographical expansions, which would probably reduce the constraining power on the lowest-order cosmographical parameters, as e.g. $q_0$. This is easy to understand if we think of cosmography as what it indeed consists in, i.e. Taylor expansions of the scale factor and derived quantities around the current time. Moreover, the redshift range of these data points already covers the fraction of the cosmological history we are mainly interested in, including the deceleration-acceleration transition point.

\subsection{Data on $E(z)$ from the Pantheon+MCT SnIa compilation}\label{sec:2p1}

%%%%%%%%%%%%%%%%%%%%%%%%%%%%%%%%%%%%%%%%%%%%%%%%%%%%%%%%%%%%%%
\begin{table*}
\begin{center}
\begin{scriptsize}
\begin{tabular}{|c|c|cccccc|}

\multicolumn{1}{c}{$z_i$} & \multicolumn{1}{c}{$E(z_i)$} & \multicolumn{6}{c}{Correlation matrix}
\\
\hline

$0.07$& $0.997\pm0.023$ & $1.00$  &   &   &   &   &  \\
 $0.20$& $1.111\pm0.020$&  $0.39$ &  $1.00$ &   &   &   & \\
 $0.35$& $1.128\pm0.037$ & $0.53$  & $-0.14$  & $1.00$  &   &   & \\
 $0.55$& $1.364\pm0.063$ &  $0.37$ &  $0.37$ & $-0.16$  & $1.00$  &  & \\
 $0.90$& $1.52\pm0.12$&  $0.01$ &  $-0.08$ &  $0.17$ &  $-0.39$ & $1.00$ &\\
 $1.50$& $2.78\pm0.59$ & $-0.03$ & $-0.07$ & $-0.07$ & $0.13$ & $-0.16$ & 1.00\\\hline
\end{tabular}
\end{scriptsize}
\caption{Data on the Hubble rate $E(z_i)$ and corresponding correlation matrix from the Pantheon+MCT SnIa compilation \cite{Riess2017,Pantheon}. The correlation matrix is of course symmetric, so we only write the elements of its lower triangle. See the text for details.\label{tab:PantheonMCTtab}}
\end{center}
\end{table*}

%%%%%%%%%%%%%%%%%%%%%%%%%%%%%%%%%%%%%%%%%%%%%%%%%%%%%%%%%%%%%

In this work we use the Hubble rate data points, i.e. $E(z_i)=H(z_i)/H_0$ with $H_0=H(z=0)$, provided in \cite{Riess2017} for six different redshifts in the range $z\in[0.07,1.5]$. They compress very effectively the information about the $1048$ SnIa at $z<1.5$ that take part of the Pantheon compilation \cite{Pantheon} (which includes the 740 SnIa of the joint light-curve analysis (JLA) sample compiled in \cite{BetouleJLA}), and the 15 SnIa at $z>1$ of the CANDELS and CLASH Multy-Cycle Treasury programs obtained by the HST, 9 of which are at $1.5<z<2.3$. Riess et al. converted in \cite{Riess2017} the raw SnIa measurements into data on $E(z)$ by parametrizing $E^{-1}(z)$ at those six redshifts $z_i$. The integral over $E^{-1}$ that defines the luminosity distance is then obtained by interpolating between $z_i$ with cubic Hermite polynomials. Finally, the overall constant $H_0$ is marginalized away along with the absolute supernovae magnitude, see \cite{Riess2017} for further details. The corresponding values of $E^{-1}(z_i)$ are Gaussian in very good approximation and are shown in Table 6 of \cite{Riess2017}, together with the corresponding correlation matrix. We present their inverse, $E(z_i)$, and the correlation matrix in Table 1 for completeness and because we will use the $E(z)$-data in the reconstruction of the Hubble rate. This will allow us to compute the weights of the WFR method exactly in this case. Notice that the correlation matrix for the $E(z)$-data is very similar to the one that contains the correlations between the $E^{-1}(z_i)$-values. The firsts five points are almost perfectly Gaussian too. In contrast, $E(z=1.5)$ is not normal-distributed at such good level, see the last plot in Fig. 1. The best-fit value reads $E(1.5)=2.67^{+0.83}_{-0.52}$. Nevertheless, we have opted to fit a Gaussian to the exact histogram as a first approximation, obtaining $E(1.5)=2.78\pm 0.59$\footnote{Other authors, as those of Refs. \cite{Haridasu2018,Pinho2018} just symmetrize the upper and lower bounds of $E(1.5)$ provided in \cite{Riess2017}, without adapting its central value. This yields $E(1.5)=2.67\pm 0.68$. No important differences in the final results are obtained when this value is used instead of ours due to the reasons exposed above in the text.}. This works quite well, since as we already showed in \cite{GomezAmendola2018}, the relative uncertainty of $E(z=1.5)$ is considerably larger than the other five data points and hence its impact on the final shape of the reconstructed functions is much lower. In addition, it is easier and more practical to deal with a multivariate Gaussian distribution, rather than considering the small departures from it, especially when their impact is so small, as in the case under study. As we will see in Sect. \ref{sec:WFR}, this is because in this way we can derive analytically also the constraints on the coefficients of the reconstructed Hubble rate, which are also Gaussian-distributed due to the fact that the latter is built linear in the parameters. This allows us to save valuable computational time.

%%%%%%%%%%%%%%%%%%%%%%%%%%%%%%%%%%%%%%%%%%%%%%%%%%%%%%%%%%%%%
\begin{figure}
\begin{center}
\includegraphics[scale=0.59]{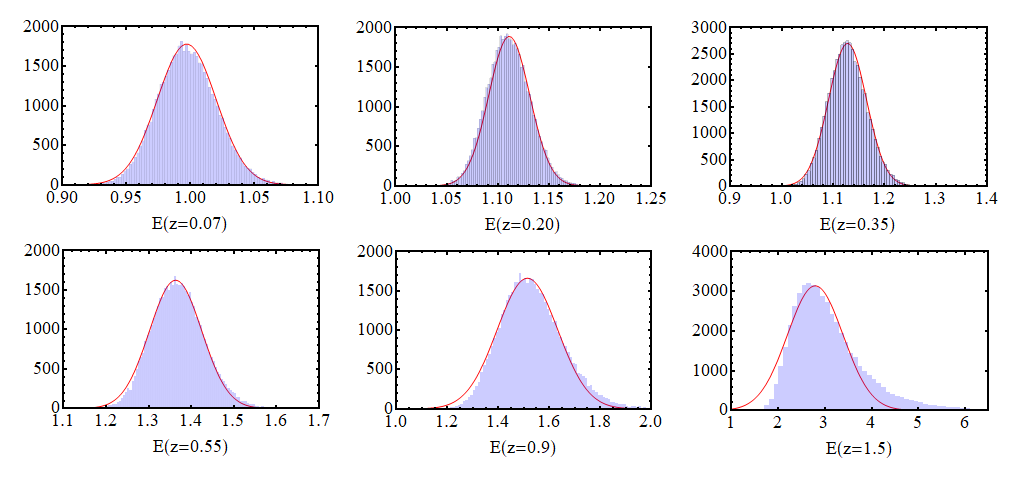}
\caption{Histograms for the six values of $E(z_i)$ derived from the Pantheon+MCT SnIa compilation. They have been obtained, together with the corresponding covariance matrix, by inverting the values of $E^{-1}(z_i)$ provided in \cite{Riess2017} with a Monte Carlo routine \cite{Metropolis,Hastings}, with which we have generated a Markov chain of $5\cdot 10^4$ points. We also superimpose the fitted Gaussians (in red) in order to show that the exact distributions are normal in very good approximation for the first five redshifts, and in lesser extent for the sixth one. See the comments in the main text of Sect. \ref{sec:2p1}.}
\end{center}
\end{figure}

%%%%%%%%%%%%%%%%%%%%%%%%%%%%%%%%%%%%%%%%%%%%%%%%%%%%%%%%%%%%%%

It is important to remark that these values on $E(z_i)$ have been obtained by assuming a flat Universe and the Cosmological Principle, and thus are model-dependent in this sense, cf. \cite{Riess2017}. The homogeneity and isotropy of the Universe at large scales are features exceedingly sustained by radiation backgrounds as CMB observations, and counts of sources observed at wavelengths ranging from radio to gamma rays. We know moreover that the flatness assumption is quite reasonable if our main aim is to use these data points to reconstruct $E(z)$ around the current time. Note e.g. that the TT+lowP+lensing+BAO analysis carried out by the Planck Collaboration (2018) \cite{Planck2018} leads to a value of $\Omega_k^{(0)}=0.0007\pm0.0019$ at $1\sigma$ c.l., which is fully compatible with the flat Universe scenario; or the analysis by Ooba, Ratra and Sugiyama \cite{Ooba2018}, which in this case favors a closed Universe, although the central value for $\Omega_k^{(0)}$ is still low, around $-0.006$ when the model is confronted to the TT,TE,EE+lowP+lensing+BAO data. One can easily check that the relative change on $H(z)$ caused by these tiny deviations from flatness is really small, being around $0.3\%$ at most (at $1\sigma$ c.l.) in the redshift range $0\leq z\leq 2$. This is of course much smaller than the relative uncertainties of our data points and also than the one of the reconstructed functions (see e.g. Figs. 3-4). Thus, given the sensitivity of the data we are dealing with, the assumption of a flat Universe has a derisory impact on our results. Moreover, it also allows us to break the existing strong degeneracy between the current values of the jerk and the snap parameters and $\Omega_k^{(0)}$ \cite{Visser2004,Visser2005,DunsbyLuongo2016}, and this is of course crucial to obtain tighter constraints on these cosmographical quantities.   

We also want to mention that the Hubble rate data of Table 1 have been obtained without considering the potential time evolution of the SnIa absolute luminosity, hence sticking to the standard approach in the literature. The authors of \cite{Tutusaus2017} interestingly showed that when this assumption is not taken for granted a decelerated low-redshift power law model of the type $a(t)\sim t^n$ (with $n<1$) is able to fit the low-redshift background data as well as, or even slightly better, than the $\Lambda$CDM. Riess et al. argued in \cite{Riess2017}, though, that when SnIa data at $z>1.5$ are included in the analysis the $\Lambda$CDM is $\sim 60$ times more probable than a marginally accelerating power-law cosmology with $n=1.04$. They conclude that there is no motivation for including the potential redshift-dependence of the intrinsic SnIa luminosity based on astrophysical or empirical considerations.

\subsection{Data on $E(z)$ from cosmic chronometers and BAOs}\label{sec:2p2}

Spectroscopic dating techniques of passively–evolving galaxies, i.e. galaxies with old stellar populations and low star formation rates, have become a good tool to obtain observational values of the Hubble function at redshifts $z\lesssim  2$ (see the work by Jim\'enez and Loeb \cite{JimenezLoeb2002} and also the references in Table 2). The measurements of CCH listed in Table 2 have been obtained from galaxies located at different angles in the sky. Under the coverage of the Cosmological Principle the dependence of the CCH data on the angle and location of the measured galaxies is removed, and therefore the $H$'s become just functions of the redshift. These measurements are independent of the Cepheid distance scale and do not rely on any particular cosmological model, although are subject to other sources of systematic uncertainties, as to the ones associated to the modeling of stellar ages, see e.g. \cite{Moresco2012,Moresco2016}, which is carried out through the so-called stellar population synthesis (SPS) techniques, and also to a possible contamination due to the presence of young stellar components in such quiescent galaxies \cite{Corredoira2017,Corredoira2018,Moresco2018}. Given a pair of ensembles of passively-evolving galaxies at two different redshifts it is possible to infer $dz/dt$ from observations under the assumption of a concrete SPS model and compute $H(z)=-(1+z)^{-1}dz/dt$. Thus, cosmic chronometers allow us to directly obtain the value of the Hubble function at different redshifts, contrary to other probes which do not directly measure $H(z)$, but integrated quantities as e.g. luminosity distances. In Table 2 we list the CCH data points used in our analyses, including their corresponding uncertainties $\sigma_i$. We point out that we have used a diagonal covariance matrix for these data, i.e. $C_{ij}=\sigma_i^2\delta_{ij}$. Moreover, we have not directly used in this study the original data points provided in the references of Table 2, $H^{\rm ori}(z_i)$'s, but the corresponding processed values, $H^{\rm pro}(z_i)$'s, obtained upon correcting the former in order to include the systematic effects mentioned before. Namely, for the data of Refs. \cite{Moresco2012,Moresco2016}, where the values of $H^{\rm ori}(z_i)$ obtained from the two alternative SPS models of \cite{BC03} and \cite{MaStro} are provided (from now on we will refer to them as BC03 and MaStro, respectively), we have opted to compute the corresponding processed value at each redshift as the weighted sum of the two, 
\begin{equation}\label{eq:Cor1}
H^{\rm pro}(z_i)=\frac{\sum\limits_{j=1}^{2}\frac{H^{\rm ori}_j(z_i)}{\sigma_j^2(z_i)}}{\sum\limits_{j=1}^{2}\sigma_j^{-2}(z_i)}\,.
\end{equation} 
The $\sigma_j(z_i)$'s do not refer to the uncertainties of the second column of Table 2, but to the corrected ones,
\begin{equation}\label{eq:Cor2}
\sigma_j(z_i)=\sqrt{\tilde{\sigma}_j^2(z_i)+[H^{\rm ori}_1(z_i)-H^{\rm ori}_2(z_i)]^2+[0.025H_j^{\rm ori}(z_i)]^2}\,,
\end{equation}
%
%%%%%%%%%%%%%%%%%%%%%%%%%%%%%%%%%%%%%%%%%%%%%%%%%%%%%%%%%%%%%%
\begin{table*}
\begin{center}
\begin{scriptsize}
\begin{tabular}{|cccc|}
\multicolumn{1}{c}{$z_i$} & \multicolumn{1}{c}{$H^{{\rm ori}}(z_i)$ [km$\,{\rm s^{-1}Mpc^{-1}}$]} & \multicolumn{1}{c}{$H^{\rm pro}(z_i)$ [km$\,{\rm s^{-1}Mpc^{-1}}$]} &  \multicolumn{1}{c}{References}\\\hline

$0.07$ & $69.0\pm 19.6$ & $69.0\pm 19.7$ & \cite{Zhang}
\\\hline
$0.09$ & $69.0\pm 12.0$ & $69.0\pm 12.1$ & \cite{Jimenez}
\\\hline
$0.12$ & $68.6\pm 26.2$ & $68.6\pm 26.3$ & \cite{Zhang}
\\\hline
$0.17$ & $83.0\pm 8.0$ & $83.0\pm 8.3$ & \cite{Simon}
\\\hline
$0.1791$ & $75.0\pm 4.0$ & $77.8\pm 8.1$ & \cite{Moresco2012}\\
& $81.0\pm 5.0$ & &
\\\hline
$0.1993$ & $75.0\pm 5.0$ & $77.7\pm 8.7$ & \cite{Moresco2012}\\
& $81.0\pm 6.0$ &  &
\\\hline
$0.2$ & $72.9\pm 29.6$ & $72.9\pm 29.7$ & \cite{Zhang}
\\\hline
$0.27$ & $77.0\pm 14.0$ & $77.0\pm 14.1$ & \cite{Simon}
\\\hline
$0.28$ & $88.8\pm 36.6$ & $88.8\pm 36.7$ & \cite{Zhang}
\\\hline
$0.3519$ & $83.0\pm 14.0$ & $85.2 \pm 16.9$ & \cite{Moresco2012}\\
& $88.0\pm 16.0$ & &
\\\hline
$0.3802$ & $83.0\pm 13.5$ & $86.0\pm 15.6$ & \cite{Moresco2016}\\
& $89.3\pm 14.1$ &&
\\\hline
$0.4$ & $95.0\pm 17.0$ & $95.0\pm 17.2$ & \cite{Simon}
\\\hline
$0.4004$ & $77.0\pm 10.2$ & $79.8\pm 12.3$ & \cite{Moresco2016}\\
& $82.8\pm 10.6$ &&
\\\hline
$0.4247$ & $87.1\pm 11.2$ & $90.3\pm 13.6$ & \cite{Moresco2016}\\
& $93.7\pm 11.7$ &&
\\\hline
$0.4497$ & $92.8\pm 12.9$ & $96.1\pm 15.3$ & \cite{Moresco2016}\\
& $99.7\pm 13.4$ &&
\\\hline
$0.47$ & $89.0\pm 49.6$ & $89.0\pm 49.6$ & \cite{Ratsimbazafy2017}
\\\hline
$0.4783$ & $80.9\pm 9.0$ & $83.8\pm 10.8$ & \cite{Moresco2016}\\
& $86.6\pm 8.7$ &&
\\\hline
$0.48$ & $97.0\pm 62.0$ & $97.0\pm 62.0$ & \cite{Stern}
\\\hline
$0.5929$ & $104.0\pm 13.0$ & $106.7\pm 16.4$ & \cite{Moresco2012}\\
& $110.0\pm 15.0$ &&
\\\hline
$0.6797$ & $92.0\pm 8.0$ & $94.6\pm 11.9$ & \cite{Moresco2012}\\
& $98.0\pm 10.0$ & &
\\\hline
$0.7812$ & $105.0\pm 12.0$ & $96.3\pm 21.0$ & \cite{Moresco2012}\\
& $88.0\pm 11.0$ &&
\\\hline
$0.8754$ & $125.0\pm 17.0$ & $124.5\pm 17.3$ & \cite{Moresco2012}\\
& $124.0\pm 17.0$ &&
\\\hline
$0.88$ & $90.0\pm 40.0$ & $90.0\pm 40.1$ & \cite{Stern}
\\\hline
$0.9$ & $117.0\pm 23.0$ & $117.0 \pm 23.2$ & \cite{Simon}
\\\hline
$1.037$ & $154.0\pm 20.0$ & $132.5\pm 45.8$ & \cite{Moresco2012}\\
& $113.0\pm 15.0$ &&
\\\hline
$1.3$ & $168.0\pm 17.0$ & $168.0\pm 17.5$ & \cite{Simon}
\\\hline
$1.363$ & $160.0\pm 33.6$ & $160.0\pm 33.8$ &  \cite{Moresco2015}
\\\hline
$1.43$ & $177.0\pm 18.0$ & $177.0\pm 18.5$ & \cite{Simon}
\\\hline
$1.53$ & $140.0\pm 14.0$ & $140.0\pm 14.4$ & \cite{Simon}
\\\hline
$1.75$ & $202.0\pm 40.0$ & $202.0\pm 40.3$ & \cite{Simon}
\\\hline
$1.965$ & $186.5\pm 50.4$ & $186.5\pm 50.6$ & \cite{Moresco2015}\\\hline

\end{tabular}
\end{scriptsize}
\caption{Data on $H(z_i)$ obtained from CCH. See the quoted references and the text for details. Notice that we write both, the original values provided in these references ($H^{\rm ori}$) and also the processed ones ($H^{\rm pro}$), i.e. those that are obtained upon the implementation of the corrections explained in Sect. \ref{sec:2p2}. In the case of Refs. \cite{Moresco2012,Moresco2016} the authors provide the values obtained with the BC03 and MaStro SPS models. We list both here, being those at the top of the corresponding row (second column) the BC03 ones and those at the bottom the MaStro ones.\label{tab:CCHtab}}
\end{center}
\end{table*}

%%%%%%%%%%%%%%%%%%%%%%%%%%%%%%%%%%%%%%%%%%%%%%%%%%%%%%%%%%%%%%

\noindent
where $\tilde{\sigma}_j(z_i)$ for $j=1,2$ are just the original uncertainties (which {\it do} refer to those of the second column of Table 2), the second term in the square root is introduced to account for the systematic error that is due to the choice of the SPS model\footnote{In Sect. \ref{sec:results} we will also report on the results that are obtained by considering a less conservative systematic uncertainty coming from the choice of SPS model, which is given in this case by half the difference of the central values.}, and the last term accounts for the potential contamination of the passively-evolving galaxies for the presence of a young stellar component. Moresco et al. reanalyzed in \cite{Moresco2018} the data presented in \cite{Moresco2012,Moresco2016} to assess the impact of this effect and showed that the young population contamination is actually minimal and consistent with zero given the current uncertainties. They calculated that at most it would bias the determinations on $H(z_i)$ by $0.4-1\%$ (at $1\sigma$, $0.8-2.3\%$ at $2\sigma$), well below the current errors. We have been conservative, though, and added a $2.5\%$ systematic uncertainty (at $1\sigma$) not only to the values reported in \cite{Moresco2012,Moresco2016}, but also to those provided in the other references, see the Table 2. We have assumed, therefore, that the conclusions of \cite{Moresco2018} can also be extended to the rest of studies from which we have compiled the CCH data. The results of the processed $H^{\rm pro}(z_i)$'s, which incorporate the corrections of formulas (\ref{eq:Cor1}) and (\ref{eq:Cor2}), are listed in the third column of the same table. The uncertainties of the $H^{\rm pro}(z_i)$'s from \cite{Moresco2012,Moresco2016} are taken to be the greatest of the two $\sigma_j$'s in each case. For the rest of references they are given by the corresponding $\sigma(z_i)$'s. 
 
In this work we also include data on BAOs. More concretely, we consider the radial component of the anisotropic BAOs obtained from the measurement of: (i) the power spectrum and bispectrum from the Baryon Oscillation Spectroscopic Survey (BOSS) data release 12 galaxies \cite{GilMarin2017}, $H(z=0.32)r_s(z_d)=(11.55\pm 0.38)\cdot 10^{3}\,{\rm km\,s^{-1}}$ and $H(z=0.57)r_s(z_d)=(14.02\pm 0.22)\cdot 10^{3}\,{\rm km\,s^{-1}}$; (ii) the complete Sloan Digital Sky Survey (SDSS) III Ly$\alpha$-quasar auto and cross-correlation functions \cite{Bourboux2017}, $c/[H(z=2.40)r_s(z_d)]=8.94\pm 0.22$; and (iii) the SDSS-IV extended BOSS data release 14 quasar sample \cite{GilMarin2018}, $H(z=1.52)r_s(z_d)=(24.0\pm 1.8)\cdot 10^{3}\,{\rm km\,s^{-1}}$. The theoretical expression of the sound horizon at the redshift of the radiation drag $z_d$ reads, 
\begin{equation}
r_s(z_d)=\int_{z_d}^{\infty}\frac{c_s(z)}{H(z)}dz\,,
\end{equation}
where $c_s(z)$ is the sound speed in the baryon-photon plasma. $r_s(z_d)$ obviously depends on the physics at very high redshifts, i.e. at $z>z_d\sim \mathcal{O}(10^3)$, and, in particular, on the Hubble function at those epochs. This complicates in principle the cosmographic analysis, since the latter is only consistent and efficient when only data at low and intermediate redshifts are included. One way to deal with this problem is to apply a reasonable (as model-independent as possible) prior for $r_s(z_d)$ in order to re-express the BAOs constraints just in terms of the value of the Hubble function at the intermediate redshifts explored by the various surveys. The Planck Collaboration (2018) \cite{Planck2018} has found $r_s(z_d)=(147.21\pm 0.48)$ Mpc fitting the $\Lambda$CDM model to the TT+lowE CMB data, almost exactly the same value that reported two years before in \cite{Planck2015} using the TT+lowP data set, $r_s(z_d)=(147.33\pm 0.49)$ Mpc. Verde et al. showed in \cite{Verde2017}, though, that when non-standard dark radiation components are allowed to be present in the pre-recombination epoch a slightly larger value of $r_s(z_d)$ is preferred, with substantial larger relative uncertainty, $r_s(z_d)=(150\pm 5)$ Mpc. We deem this is a more model-independent estimation of the sound horizon at the drag epoch. It covers the Planck preferred value and range at $<1\sigma$ c.l., and is also compatible with the model-independent determinations provided in Ref. \cite{Heavens2014}, $r_s(z_d)=(142.8\pm 3.7)$ Mpc, and \cite{Haridasu2018}, $r_s(z_d)=(145.6\pm 5.1)$ Mpc. In contrast to the values reported in these two references, the value from \cite{Verde2017} is not extracted only from low and intermediate-redshift data, but incorporates the CMB information too. The latter is a key ingredient, of course, since the CMB anisotropies are basically fixed by the pre-recombination physics and, therefore, the latter are important to constrain $r_s(z_d)$ without setting aside the physical processes occurred before the decoupling of photons from baryons. Although the sound horizon from \cite{Verde2017} is not fully model-independent, the assumptions under which it has been obtained are relaxed enough to be considered, in practice, as so \footnote{We will also discuss the impact of the choice of the prior of $r_s(z_d)$ in Sect. \ref{sec:results}, by also providing the results obtained with the value from \cite{Heavens2014}.}. For all these reasons we are going to use it as a prior in our study to transform the BAOs information into direct constraints on $H(z_i)$. But we are actually not only interested in doing this, but also in obtaining direct constraints on the Hubble rate $E(z_i)$ from the aforesaid data on BAOs and also the CCH data points listed in Table 2. It is highly convenient to work only with data on $E(z_i)$ because, as we will see later on more in detail, the combination of data on $H(z_i)$ and $E(z_i)$ would make appear in the description of $H(z)$ some non-linearities due to the product of $H_0$ with the parameters of the linear expansion that we will use for the reconstruction of $E(z)$. These non-linearities would pose a problem because then the parameters of $E(z)$ would not be Gaussian-distributed, and the exact computation of the evidences would have to be carried out numerically, which would imply a much more expensive budget in terms of computational time. In contrast, if we only use data on $E(z_i)$ we are able to compute the exact evidences and Bayes ratios (and thus the weights of the WFR method, see Sect. \ref{sec:evidences}) in a pure analytical way, just because the data is Gaussian-distributed and $E(z)$ will be constructed linear in the parameters. We will provide the corresponding expressions in the subsequent section. Hence, according to these explanations it becomes clear that we need to also include a prior on $H_0$ (again, as model-independent as possible) in our analysis. The choice of this prior is not a straightforward task. It is very well-known that there exists a $3.5\sigma$ tension between the local determination of $H_0$ by Riess et al. \cite{RiessH02018}, $H_0=(73.48\pm 1.66)$ km$\,{\rm s^{-1}Mpc^{-1}}$, and the one found by the Planck Collaboration (2018) \cite{Planck2018} assuming the $\Lambda$CDM and using e.g. the TT+lowE CMB data, $H_0=(66.88\pm 0.92)$ km$\,{\rm s^{-1}Mpc^{-1}}$. Whether this tension is caused by some kind of (unaccounted) systematic error affecting the data or due to the need of new physics beyond the standard model is still unknown. Other works that make use of the cosmic distance ladder find very similar results to the one reported by \cite{RiessH02018}, see e.g. \cite{Cardona2017,JangLee2017,Zhang2017,Follin2018}, and are also consistent with preceding studies, as e.g. \cite{Riess2011,RiessH02016}. Any important systematic error has been neither found in Planck's determination. Addison et al. showed in \cite{Addison2018} that independent analyses from Planck using alternative high-redshift data also lead to results in non-negligible tension with the local value of $H_0$. For instance, using the constraints on the primordial abundance of deuterium and galaxy and Ly$\alpha$ forest BAOs data they found $H_0=(66.98\pm 1.18)$ km$\,{\rm s^{-1}Mpc^{-1}}$ in the context of the $\Lambda$CDM. Assuming the same model, Bonvin et al. found $H_0=(71.9^{+2.4}_{-3.0})$ km$\,{\rm s^{-1}Mpc^{-1}}$ by analyzing the gravitational time delay of the light rays coming from the three multiply imaged quasar system HE $0435-1223$ \cite{Bonvin2017}, and Birrer et al. $H_0=(68.8^{+5.4}_{-5.1})$ km$\,{\rm s^{-1}Mpc^{-1}}$ from the  doubly imaged quasar SDSS 1206+4332 \cite{Birrer2018}. In these cases there is no tension with the local determination. Recently, it has also been possible to measure the Hubble parameter using the gravitational wave signal of the neutron star merger GW17081716 and its electromagnetic counterpart \cite{Abbott2017,Guidorzi2017}, providing high values of $H_0$, but still with very large uncertainties ($70^{+12}_{-8}$ and $75.5^{+11.6}_{-9.6}$ km$\,{\rm s^{-1}Mpc^{-1}}$, respectively). They are completely independent from the underlying cosmology and the cosmic distance ladder. The claimed reduction of the degeneracy between the source distance and the weakly constrained viewing angle has allowed to considerably reduce also the uncertainty of $H_0$, yielding $68.9^{+4.7}_{-4.6}$ km$\,{\rm s^{-1}Mpc^{-1}}$ \cite{Hotokezaka2018}. The central value is now more compatible with the local determinations, although some criticisms to this new estimation have been also drawn \cite{DadoDar2018}. It is also worth to mention the cosmology-independent analyses carried out in \cite{YuRatraWang2017,GomezAmendola2018,Feeney2018,Haridasu2018,Lemos2018,Macaulay2018,LiDuXu2019}, where use is made of different combinations of cosmological low and intermediate-redshift data involving SnIa, CCH and BAOs. Values of $H_0$ lying in the range $\sim 67-68.5$ km$\,{\rm s^{-1}Mpc^{-1}}$ are obtained, see these references for details.  Some other authors, as those from \cite{Marra2013} and, more recently, those from \cite{WuHuterer,CamarenaMarra,Bengaly2018}, have studied the impact of the cosmic variance on the local determination of $H_0$, and conclude that although it might certainly play a role its effect is unable to explain the whole discrepancy between the HST and Planck's values. Romano explains in his paper \cite{Romano2018} that the use by Riess et al. of the 2M++ density field map (which covers redshifts $z\leq 0.06$) to compute peculiar velocity flows could be biasing their results, since there is evidence of the existence of local radial inhomogeneities extending in different directions up to a redshift of about $0.07$ \cite{Keenan2013}, and according to \cite{Romano2018} the $40\%$ of the Cepheids used in \cite{RiessH02018} would be affected. Moreover, Shanks, Hogarth \& Metcalfe claim in \cite{Shanks2018a} that the GAIA parallax distances of Milky Way Cepheids employed in \cite{RiessH02018} in the first step of the cosmic distance ladder may be underestimated a $\sim 7-18\%$, and this would also produce an important decrease of their measured value of $H_0$. The discussion on the validity of these arguments is, though, still open and intense \cite{RiessProb,Shanks2018b}. In view of the large dispersion of values found in the vast literature, we opt to adopt in this work the following prior in our main analyses: $H_0=(70\pm 5)$ km$\,{\rm s^{-1}Mpc^{-1}}$, which basically covers the values of interest without relying exclusively on the low or the high parameter region. We have also studied, though, the impact of the (less conservative) priors provided in \cite{RiessH02018,Planck2018} and checked that they lead to fully compatible results for the shape of the cosmographical functions and, in particular, for the value and the uncertainty of $q_0$. Finally, we consider a correlation coefficient $\rho=-0.56$ between $H_0$ and $r_s(z_d)$ in the Gaussian prior of the main analyses, inspired by the $\Lambda$CDM Planck's constraints. We have also explored other values around -0.6. The results are kept consistent too. Using this two-dimensional prior we can convert the CCH values of $H(z_i)$ and the BAOs data into direct constraints on $E(z_i)$, which can later on be employed together with the SnIa data listed in Table 1 to perform the cosmographical analysis with the WFR method, using only data on the Hubble rate at different redshifts. We have checked that the distribution of the processed CCH and BAOs data on $E(z_i)$ is very well approximated by a multivariate Gaussian, which is crucial to compute analytically the evidence associated to the various cosmographical expansions of $E(z)$, see Sect. \ref{sec:evidences}. As expected, although the original CCH and BAOs data are uncorrelated, some correlations appear between the corresponding transformed values of $E(z_i)$ due to the use of the common prior on $H_0$ and $r_s(z_d)$. We have taken all these correlations into account. 

%%%%%%%%%%%%%%%%%%%%%%%%%%%%%%%%%%%%%%%%%%%%%%%%%%%%%%%%%%%%%%
%%%%%%%%%%%%%%%%%%%%%%%%%%%%%%%%%%%%%%%%%%%%%%%%%%%%%%%%%%%%%%
\section{The Weighted Function Regression method}\label{sec:WFR}

%%%%%%%%%%%%%%%%%%%%%%%%%%%%%%%%%%%%%%%%%%%%%%%%%%%%%%%%%%%%%%
%%%%%%%%%%%%%%%%%%%%%%%%%%%%%%%%%%%%%%%%%%%%%%%%%%%%%%%%%%%%%%
\subsection{Cosmography through $E(z)$: deceleration, jerk, and snap parameters}\label{sec:cosmography}

Assuming (as we do in this paper) that the Universe is flat, homogeneous and isotropic, the square of the space-time element interval can be written in the usual Friedmann-Lema\^itre-Robertson-Walker (FLRW) form, 
\begin{equation}
ds^2=dt^2-a^2(t)d\vec{x}^2\,,
\end{equation}
with $a(t)$ being the scale factor and $x_i$ with $i=1,2,3$ the comoving space coordinates. The former can be Taylor-expanded around the current time $t_0$, 
\begin{equation}\label{eq:a1}
a(t)=1+\sum_{n=1}^{\infty}\frac{1}{n!}\frac{d^na}{dt^n}\Bigr\vert_{t=t_0}(t-t_0)^n\,, 
\end{equation}
where we have set $a(t_0)=1$. It can also be written in terms of the current values of the various cosmographical functions. Up to the fourth order in $t$, the expansion involves the Hubble and deceleration parameters (see Eq. (\ref{eq:decelerationParam}) and below), as well as the jerk and the snap, which read respectively \cite{Visser2004,Visser2005},
\begin{equation}\label{eq:jerksnap1}
j=\frac{1}{aH^3}\frac{d^3a}{dt^3}\,,\qquad s=\frac{1}{aH^4}\frac{d^4a}{dt^4}\,.
\end{equation}
Upon substitution in (\ref{eq:a1}) one obtains, 
\begin{equation}\label{eq:a2}
a(t)=1+H_0(t-t_0)-\frac{1}{2}q_0H_0^2(t-t_0)^2+\frac{1}{6}j_0H_0^3(t-t_0)^3+\frac{1}{24}s_0H_0^4(t-t_0)^4+...\,,
\end{equation}\footnote{The minus sign of the third term on the right-hand side of (\ref{eq:a2}) is due to the definition of the deceleration parameter (\ref{eq:decelerationParam}). The latter is still defined as in \cite{Sandage1970}, when it was thought that the Universe was currently decelerating, i.e. $\ddot{a}(t_0)<0$, due to the dominance of the non-relativistic matter. The minus sign of (\ref{eq:decelerationParam}) made $q_0$ to be positive. After the works \cite{SNIaRiess,SNIaPerl} there is probably no {\it raison d'\^{e}tre} for this minus sign, but nevertheless it has been preserved in the definition, as originally.}
with the subscript $0$ referring to present-day quantities. As we have shown in Sect. \ref{sec:Data}, we will deal with data on the Hubble rate in order to extract the cosmographical information. Thus, it is better to directly express it in the cosmographical form (instead of working with the expansion of the scale factor), and as a function of the redshift (instead of the cosmic time), since the former is the physical variable with which we make contact with observations. The Taylor series of $H(z)$ around $z=0$ is just given by
\begin{equation}\label{eq:TaylorH}
H(z)=H_0+\sum_{n=1}^{\infty}\frac{1}{n!}\frac{d^nH}{dz^n}\Bigr\vert_{z=0}z^n\,.
\end{equation}
In order to write this expansion in terms of the cosmographical parameters we firstly need to express the cosmographical functions (\ref{eq:decelerationParam}) and (\ref{eq:jerksnap1}) in terms of the derivatives of $H(z)$ with respect to the redshift $z$. Making use of $d/dt=-(1+z)H(z)d/dz$ and after a little bit of algebra we obtain the following relations,
\begin{equation}\label{eq:q(z)1}
q(z)=-1+\frac{1+z}{H(z)}\frac{dH}{dz}\,,
\end{equation}
\begin{equation}\label{eq:j(z)1}
j(z)=q^2(z)+\frac{(1+z)^2}{H(z)}\frac{d^2H}{dz^2}\,,
\end{equation}
\begin{equation}\label{eq:s(z)1}
s(z)=3[q^2(z)+q^3(z)-j(z)]-4q(z)j(z)-\frac{(1+z)^3}{H(z)}\frac{d^3H}{dz^3}\,.
\end{equation}
We can isolate the derivatives of the Hubble function from these expressions and use them in (\ref{eq:TaylorH}). Dividing the result by $H_0$ one finally obtains the expansion of the Hubble rate in terms of the current values of the cosmographical functions, as desired,
\begin{eqnarray}\label{eq:EexpZ}
E(z)=1&+&(1+q_0)z+\frac{1}{2}(j_0-q_0^2)z^2+\frac{1}{6}(3q_0^3+3q_0^2-3j_0-4q_0j_0-s_0)z^3+...
\end{eqnarray}
As it was already reported in \cite{CattoenVisser2007a,CattoenVisser2007b} this Taylor series (which is built in $z$, around $z=0$) converges for $|z|<1$ and therefore we cannot expect it to describe the correct physical behavior at redshifts larger than one. One possible way to solve this problem is to apply a shift in the pivoting redshift in order to increase the radius of convergence. Another viable solution consists in constructing the Taylor series in the variable $y=1-a=z/(1+z)$ instead of directly $z$, as it is also suggested in \cite{CattoenVisser2007a,CattoenVisser2007b}, see therein for further details. Let us just mention that in this case the series converges for $|y|<1$ or, equivalently, in the redshift range $z\in (-0.5,+\infty)$. The Taylor series of the Hubble rate around $y=0$ reads as follows,
\begin{figure}
\begin{center}
\includegraphics[scale=0.6]{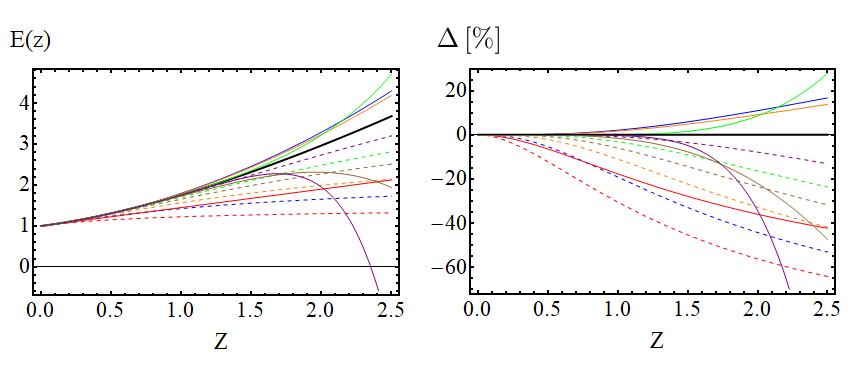}
\caption{{\it Left plot:} Hubble rate $E(z)$ for the $\Lambda$CDM model using $\Omega_m^{(0)}=0.3$ (in black, cf. Eq. \eqref{eq:ECPL} with $(w_0,w_1)=(-1,0)$). The other curves correspond to the associated truncated expansions $E_t(z)$ of the Taylor series of $E(z)$ up to order 1, 2, 3, 4, 5, 7 (in red, blue, orange, brown, green, purple, respectively). The continuous curves are obained from (\ref{eq:EexpZ}), whereas the dashed ones are obtained from (\ref{eq:EexpY}); {\it Right plot:} Relative differences of the truncated expansions with respect to the $\Lambda$CDM curve expressed in [\%], i.e. $\Delta = 100\cdot(E_t(z)/E(z)-1)$. See comments in the text.}
\end{center}
\end{figure}
\begin{equation}
E(y)=1+\sum_{n=1}^{\infty}\frac{1}{n!}\frac{d^nE}{dy^n}\Bigr\vert_{y=0}y^n\,,
\end{equation}
and can be also written as
\begin{eqnarray}\label{eq:EexpY}
E(z)=&& 1+(1+q_0)\frac{z}{1+z}+(j_0-q_0^2+2+2q_0)\frac{z^2}{2(1+z)^2}\\
&&+(6+6q_0-3q_0^2+3q_0^3+3j_0-4j_0q_0-s_0)\frac{z^3}{6(1+z)^3}+...\nonumber
\end{eqnarray}
One can check that Taylor-expanding $(1+z)^{-1}=1-z+z^2-z^3+\mathcal{O}(z^4)$ in all the terms of the last expression one retrieves (\ref{eq:EexpZ}), so both formulas are consistent. As mentioned before, (\ref{eq:EexpY}) is {\it a priori} more appropriate than (\ref{eq:EexpZ}) because the former solves the formal convergence problem discussed above. This fact is usually employed by many authors in the literature as an argument to justify the use of the truncated expansions of the Taylor series in $y$ in front of the ones in $z$ in the cosmographical fitting analyses that make use of data at $z>1$. Care should be taken, though, since the convergence of the full Taylor series only ensures the better fitting performance of the truncated expansions for high enough truncation orders. This better behavior may not happen for lower ones. Actually, this is precisely the case concerning the Taylor series (\ref{eq:EexpY}). In order to see this we have presented Fig. 2. In the plot on the left we draw the curves of the Hubble rate for the $\Lambda$CDM model and its associated truncated expansions of the Taylor series (\ref{eq:EexpZ}) and (\ref{eq:EexpY}) (cf. the caption of Fig. 2 for details). The plot on the right shows the relative differences between the former and the latter. The convergent nature of (\ref{eq:EexpY}) is palpable from them. When one adds more and more terms of the Taylor series, the truncated expansions (dashed curves) tend to be closer to the exact function (drawn in black) and, hence, the relative differences decrease. Conversely, adding more terms to the truncated expansions obtained from (\ref{eq:EexpZ}) does not lead to a better description of the exact function at redshifts $z>1$ (compare, e.g. the continuous curve in green corresponding to the 5th order expansion with the 7th order one, in purple). Nevertheless, it is of utmost importance to notice that the truncated expansions up to order 5 from (\ref{eq:EexpZ}) are closer to the exact Hubble rate than those from (\ref{eq:EexpY}) at redshifts $z<2.5$, and this is so even more conspicuously at redshifts $z<1$, in the region at which the vast majority  (roughly the $80\%$) of the data points lie. Therefore, if we want to perform a cosmographical analysis with low and intermediate-redshift data, cutting the Taylor series at order 5 or lower, then it is preferable to use (\ref{eq:EexpZ}) instead of (\ref{eq:EexpY}), contrary to what it is usually advocated in other works in the literature (see the list of references provided in the Introduction). As we will explain later on, in Sect. \ref{sec:results}, the data sets described in Sect. \ref{sec:Data} do not require a high degree of complexity of the fitting expansions. They prefer those with truncation orders 2-4, so the expansions based on (\ref{eq:EexpZ}) will be preferred to those based on (\ref{eq:EexpY})\footnote{In this illustrative discussion on the performance of (\ref{eq:EexpZ}) and (\ref{eq:EexpY}) we have assumed, for simplicity, that the $\Lambda$CDM with $\Omega_m^{(0)}=0.3$ is the model that rules the Universe. This could be of course, just an approximation, but we deem it is a quite valid one, since we do not expect large deviations from the $\Lambda$CDM affecting the shape of $E(z)$ at low and intermediate redshifts.}. We will study all this applying rigorous Bayesian tools in the subsequent sections. We will analyze the performance of the two expansion types in an explicit way. Notice that we can use in principle some truncated forms of (\ref{eq:EexpZ}) and (\ref{eq:EexpY}) to extract important kinematic information about the Cosmos' expansion without making any assumption about the matter-energy sources of the gravitational field equations nor the gravitational theory itself, i.e. without assuming anything about the ultimate cause of the Universe's dynamics. We cannot only extract the values of $q_0$, $j_0$ and $s_0$, but we can also reconstruct $q(z)$, $j(z)$ and $s(z)$ using (\ref{eq:q(z)1})-(\ref{eq:s(z)1}). These truncated series are, though, not free from problems, as we will duly explain in the next subsection. We will mitigate these problems in the context of the WFR method (see Sect. \ref{sec:WFRdes} for details). In Sect. \ref{sec:results} we will show that the reconstructed shapes of $E(z)$ obtained using (\ref{eq:EexpZ}) and (\ref{eq:EexpY}) and the WFR method are fully consistent. Converseley, the situation for $q(z)$ is different. Although the central values derived with (\ref{eq:EexpZ}) and (\ref{eq:EexpY}) are compatible, the corresponding uncertainties change quite a lot depending on the parametrization employed in the analysis, and this is of course related with the discussion presented above.

In this work we will also apply the WFR method using expansions of $q(z)$ instead of expansions of the Hubble rate. More concretely, we will explore the following two possibilities,
\begin{equation}\label{eq:ExpqZ}
q(z)=q_0+q_1z+q_2z^2+...
\end{equation}
and
\begin{equation}\label{eq:ExpqY}
q(z)=q_0+\frac{q_1z}{1+z}+\frac{q_2z^2}{(1+z)^2}+...
\end{equation}
to reconstruct the deceleration parameter. We will show that we find consistent reconstructed shapes for $q(z)$. One can also rewrite (\ref{eq:j(z)1}) in terms of only $q(z)$, 
\begin{equation}
j(z)=2q^2(z)+q(z)+(1+z)\frac{dq}{dz}\,.
\end{equation}
This formula eases the direct reconstruction of the jerk from (\ref{eq:ExpqZ}) and (\ref{eq:ExpqY}). 

In Sect. \ref{sec:results} we provide all the details about the various reconstructions we have carried out in this work, and provide the constraints for $q_0$ and $j_0$ derived from them.

%%%%%%%%%%%%%%%%%%%%%%%%%%%%%%%%%%%%%%%%%%%%%%%%%%%%%%%%%%%%%%
%%%%%%%%%%%%%%%%%%%%%%%%%%%%%%%%%%%%%%%%%%%%%%%%%%%%%%%%%%%%%%

\subsection{Estimation of $q_0$ in some model and parametrization-dependent scenarios, and the need of an improved cosmographical approach}\label{sec:Motivation}
We dedicate this section to motivate the need of improving those works in the literature that obtain constraints on the cosmographical functions in the framework of concrete cosmological models, using particular parametrizations of the cosmographical functions, or even truncated cosmographical series in which the maximum order of the expansion is chosen in an {\it ad hoc} way in more or lesser extent. All these approaches are perfectly licit, of course, but one cannot claim to extract model-independent information about the Universe's kinematics from them. As an example, we will explicitly derive the constraints for $q_0$ that are obtained in some of these scenarios just to show that both, the central values and, more conspicuously, their corresponding uncertainties, are very sensitive to the particular framework chosen to carry out the analysis. This means that the results and conclusions derived from these studies can be in some cases partially or completely biased. Thus, we are forced to search for an alternative approach that proves capable of reducing the existing degree of subjectivity. Section \ref{sec:WFRdes} will be devoted to the description of one of such methods, the WFR.
%%%%%%%%%%%%%%%%%%%%%%%%%%%%%%%%%%%%%%%%%%%%%%%%%%%%%%%%%%%%%%
\begin{table*}
\begin{center}
\begin{scriptsize}
\begin{tabular}{|c|c|c|c|c|c|c|}

\multicolumn{1}{c}{Model} & \multicolumn{1}{c}{$\Omega_m^{(0)}$} & \multicolumn{1}{c}{$H_0$ [km$\,{\rm s^{-1}Mpc^{-1}}$]} & \multicolumn{1}{c}{$w_0$} & \multicolumn{1}{c}{$w_1$} &  \multicolumn{1}{c}{$q_0$} & \multicolumn{1}{c}{$\chi^2_{\rm min}$}

\\\hline 
$\Lambda$CDM & $0.295\pm0.021$ & $70.45\pm 2.36$  & $-1$ & $0$  & $-0.554\pm 0.032$ & $15.74$ \\\hline 
XCDM & $0.306\pm0.051$ & $70.31\pm 2.42$  & $-1.03\pm 0.15$ & $0$  & $-0.578\pm 0.099$ & $15.74$ \\ \hline
CPL & $0.301\pm0.104$ & $70.34\pm 2.47$  & $-1.04\pm 0.16$ & $0.10\pm 1.78$ & $-0.494\pm 0.195$ & $15.74$  
\\\hline
\end{tabular}
\end{scriptsize}
\caption{Fitting results obtained for the $\Lambda$CDM model and the XCDM and CPL parametrizations of the DE EoS (see Eqs. \eqref{eq:densities}-\eqref{eq:q0models}, and the associated comments in the text), by using the Pantheon+MCT SnIa and the CCH data described in Sect. \ref{sec:Data}. The derived deceleration parameter $q_0$ and the minimum value of the $\chi^2$-function for each model are also provided.\label{tab:fitmodels}}
\end{center}
\end{table*}
%%%%%%%%%%%%%%%%%%%%%%%%%%%%%%%%%%%%%%%%%%%%%%%%%%%%%%%%%%%%%%  
We start now analyzing some cosmological models in standard GR, considering the Cosmological Principle and a flat Universe. In this framework it is possible to write the deceleration parameter in terms of the energy densities and pressures of the various species that fill the Universe by using (\ref{eq:q(z)1}) together with the Friedmann and energy conservation equations,
\begin{equation}\label{eq:q(z)rhop}
q(z)=-1+\frac{3}{2}\frac{\sum\limits_{i}[\rho_i(z)+p_i(z)]}{\sum\limits_{i}\rho_i(z)}\,,
\end{equation}
where the subscript $i$ labels all the matter-energy components. Let us focus now in the late-time expansion, when the radiation energy density is negligible versus the non-relativistic matter one. If the latter and the DE are self-conserved then
\begin{equation}\label{eq:densities}
\rho_m(z)=\rho_m^{(0)}(1+z)^3\quad ; \quad \rho_{D}(z)=\rho_D^{(0)}e^{3\int_{0}^{z}\frac{1+w(\tilde{z})}{1+\tilde{z}}d\tilde{z}}\,.
\end{equation}
The evolution of the DE density is thus specified by the DE EoS parameter $w(z)=p_D(z)/\rho_D(z)$, and {\it vice versa}. As the pressure of the matter component is negligible versus its energy density, one can write the deceleration parameter in the case under study only in terms of $w(z)$ by plugging (\ref{eq:densities}) into (\ref{eq:q(z)rhop}). There is thus a one-to-one correspondence between $q(z)$ and the EoS parameter of the self-conserved DE, once we fix the values of the current energy densities $\rho_m^{(0)}$ and $\rho_D^{(0)}$. We analyze here three scenarios: (i) the $\Lambda$CDM model, in which $w(z)=-1$ $\forall{z}$ and the DE density remains constant throughout all the cosmic expansion, see e.g. \cite{DEbook} and references therein; (ii) the XCDM (also known as $w$CDM) parametrization of the DE EoS parameter \cite{XCDM}, in which $w(z)=w_0$, with $w_0$ being a constant that can acquire both, quintessence ($w_0>-1$) or phantom-like ($w_0<-1$) values; and (iii) the CPL parametrization \cite{CPL,Linder2003,Linder2004}, the next-to-leading order correction of the XCDM, in which $w(z)=w_0+w_1z/(1+z)$. The EoS parameter has in the latter case some evolution and could (at least, in principle) pass through the phantom divide. We choose these models basically because of their simplicity and also because they have a different number of free parameters. The XCDM and CPL have one and two more parameters than the concordance model, respectively. In the CPL parametrization the square of the Hubble rate reads, 
\begin{equation}\label{eq:ECPL}
E^2(z)=\Omega_m^{(0)}(1+z)^3+(1-\Omega_m^{(0)})e^{-\frac{3w_1 z}{(1+z)}}(1+z)^{3(1+w_0+w_1)}\,,
\end{equation}
with $\Omega_m^{(0)}=\rho_m^{(0)}/(\rho_m^{(0)}+\rho_{D}^{(0)})$ being the matter density parameter. For the $\Lambda$CDM and the XCDM the corresponding expressions are obtained straightforwardly, by just setting in (\ref{eq:ECPL}) $(w_0,w_1)=(-1,0)$ in the first case, and $w_1=0$ in the second one. Using these formulas we can confront the three models to the Pantheon+MCT and the CCH data (cf. Tables 1 and 2, respectively, and the comments in Sect. \ref{sec:Data}). The results are listed in Table 3, where we also show the value of $q_0$ that is obtained for each of the models under study. For the XCDM and CPL parametrizations the theoretical expression of the deceleration parameter reads,
\begin{equation}\label{eq:q0models}
q_0=-1+\frac{3}{2}\left[1+w_0(1-\Omega_m^{(0)})\right]\,,
\end{equation}
whereas for the $\Lambda$CDM we have to set $w_0=-1$ in this formula. Notice that the values that are obtained for this kinematic quantity are compatible in the three models, and hence fully consistent (cf. the penultimate column of Table 3). Nevertheless, the uncertainties are quite different in magnitude, being in the XCDM (CPL) a factor $\sim 3$ ($\sim 6$) larger than in the $\Lambda$CDM. The reason is obvious, in the XCDM (CPL) we have one (two) more free parameter(s) than in the concordance model, so the constraints that are obtained from the data for the various fitting parameters and derived quantities are weaker for the former models. But then, which is the level of evidence at which we can state that $q_0<0$, i.e. in favor of the current positive-accelerated phase of the Universe? In the $\Lambda$CDM the value of $q_0$ is $\sim 17\sigma$ away from $q_0=0$, in the XCDM such distance is of roughly $6 \sigma$, and in the CPL it is ``only'' of $\sim 2.5\sigma$, so the differences are not precisely small. The minimum values of the $\chi^2$-function obtained in the $\Lambda$CDM, XCDM and CPL (cf. again Table 3, last column), tell us that these models are able to fit equally well the data, so we have to use some method to penalize the use of extra parameters and see which is the most favored scenario. Whatever it is the method employed we will find e.g. that the $\Lambda$CDM is preferred over the XCDM and CPL. Although the values of $\chi^2_{\rm min}$ are the same for the three models, there is a penalization for the XCDM and CPL with respect to the $\Lambda$CDM which is caused by the addition of the free parameters $w_0$ and $(w_0,w_1)$, respectively. But still, up to what extent can we rely on the uncertainty of $q_0$ that is obtained in the framework of the $\Lambda$CDM? Are all these constraints representative of the underlying ``true'' model describing the Cosmos? It could well be not the case, since even if the $\Lambda$CDM is more preferred than the XCDM and  CPL, we have made some important assumptions that might have a non-negligible impact on our results and, more conspicuously, on the corresponding uncertainties. Apart from assuming the isotropy, homogeneity and flatness of the Universe, we have assumed that the correct theory of gravity is Einstein's GR together with the self-conservation of matter and DE, and the presence of a cosmological constant triggering the cosmic acceleration. Some of these can be considered very strong assumptions and, certainly, dispensing with them would lead to more loose constraints on $q_0$ than those obtained in the context of the $\Lambda$CDM. Thus, in order to extract more objective constraints on $q_0$ we should definitely abandon the model-dependent approach.
%
%%%%%%%%%%%%%%%%%%%%%%%%%%%%%%%%%%%%%%%%%%%%%%%%%%%%%%%%%%%%%%
\begin{table*}
\begin{center}
\begin{scriptsize}
\begin{tabular}{|c|c|c|c|c|c|}
\multicolumn{1}{c}{$q(z)$-parametrization} & \multicolumn{1}{c}{$H_0$ [km$\,{\rm s^{-1}Mpc^{-1}}$]} & \multicolumn{1}{c}{$q_0$} & \multicolumn{1}{c}{$q_1$} &  \multicolumn{1}{c}{$q_2$} & \multicolumn{1}{c}{$\chi^2_{\rm min}$}
\\\hline
$q_0$ & $72.29\pm 2.37$ & $-0.288\pm 0.036$  & $-$  & $-$ &$32.64$ \\ \hline
$q_0+q_1z$ & $70.35\pm 2.47$ & $-0.503\pm 0.063$  & $0.66\pm 0.16$ & $-$  & $16.26$ \\\hline
$q_0+q_1z/(1+z)$ & $70.55\pm 2.46$ & $-0.611\pm 0.084$  & $1.50\pm 0.36$ & $-$ & $15.74$ \\ \hline
$q_0+q_1z/(1+z)+q_2z^2/(1+z)^2$ & $70.49\pm 2.51$ & $-0.59\pm 0.20$  & $1.33\pm 1.89$ & $0.31\pm 3.24$ & $15.74$\\\hline
\end{tabular}
\end{scriptsize}
\caption{As in Table 3, but for four alternative parametrizations of $q(z)$.\label{tab:fitmodels2}}
\end{center}
\end{table*}
%%%%%%%%%%%%%%%%%%%%%%%%%%%%%%%%%%%%%%%%%%%%%%%%%%%%%%%%%%%%%%   
%

Cosmography can help us to extract model-independent constraints on the various kinematic quantities when the data employed are themselves free of model-dependencies, which unfortunately is not always the case. We have to remark, though, one important point which is usually overlooked in many works in the literature. Although the cosmographical functions (e.g. $q(z)$ or $j(z)$) can be obtained in a very model-independent way, they are {\it not} model-independent {\it per se}. For instance, by just building and fitting parametrized expressions of these cosmographical functions to the data we are not led to fully model-independent results. Given a parametrized form of $q(z)$ one can integrate (\ref{eq:q(z)1}) to obtain the associated Hubble function, and use it to compute the rest of higher order cosmographical functions as well, as the jerk (\ref{eq:j(z)1}) and the snap (\ref{eq:s(z)1}). It is also possible to relate the aforementioned parametrization of $q(z)$ with various models of DE in the standard GR scenario. Once we have $H(z)$ we can obtain the DE pressure $p_D(z)$ using the equation,
\begin{equation}\label{eq:pressureEq}
3H^2(z)-2(1+z)H(z)\frac{dH}{dz}=-8\pi G p_D(z)\,.
\end{equation}
Notice that the concrete form of the density $\rho_D(z)$ is not unequivocally determined and will exclusively depend on the way we split the conservation equation for the DE and matter,
\begin{equation}
-(1+z)\sum_i\frac{d\rho_i}{dz}+3\sum_i[\rho_i(z)+p_i(z)]=0\,,
\end{equation}
i.e. on the specific form of the source function $Q(z)$ that describes the transfer of energy from one sector to the other,
\begin{equation}
-(1+z)\frac{d\rho_m}{dz}+3\rho_m(z)=Q(z)\,,
\end{equation}
\begin{equation}
 -(1+z)\frac{d\rho_D}{dz}+3[\rho_D(z)+p_D(z)]=-Q(z)\,.
\end{equation}
In order to show this in more concrete terms, let us put a simple example in which we assume that $q(z)=q_0$, with $q_0$ being a constant. Upon integration of (\ref{eq:q(z)1}) we obtain,
\begin{equation}
H(z)=H_0(1+z)^{1+q_0}\,,
\end{equation}
and using this result in (\ref{eq:pressureEq}) we compute the DE pressure,
\begin{equation}
p_D(z)=-\frac{3H_0^2}{8\pi G}(1+z)^{2(1+q_0)}\left[1-\frac{2}{3}(1+q_0)\right]\,.
\end{equation}
If we assume that matter and DE are self-conserved, i.e. that $Q(z)=0$, we are led to the following expression for the DE density,
\begin{equation}
\rho_D(z)=\frac{3H_0^2}{8\pi G}(1+z)^{2(1+q_0)}-\rho_m^{(0)}(1+z)^3\,,
\end{equation}
\noindent
and the standard matter dilution law $\rho_m(z)=\rho_m^{(0)} (1+z)^{3}$. Different expressions for the energy densities are obtained when $Q(z)\ne 0$, and they change with the particular form of $Q(z)$. The same happens for more elaborated parametrizations of the deceleration parameter, showing in all cases that we can associate an infinite set of DE models to a given parametrization of $q(z)$. This seems to point out that the problem is somehow alleviated with respect to the cases analyzed before in which particular cosmological models were assumed, since now we can obtain constraints on cosmographical functions which are not only valid for a concrete model, but are also extensible to a whole family of them. In this sense, this approach is more model-independent. Despite this, such constraints are still very reliant on the particular parametrization chosen, as can be explicitly checked in Table 4, where we show the fitting results for four alternative parametrizations of $q(z)$. Again, the level of evidence in favor of the current positive acceleration of the Universe varies a lot depending on the particular choice of $q(z)$. It ranges from the $\sim 3\sigma$ significance of the most complex model (the one in the fourth row) to the $\sim 8\sigma$ found using e.g. $q(z)=q_0$, and the latter is in strong tension with the values obtained with the other parametrizations. This situation was already noticed by Elgar{\o}y and Multam\"{a}ki, who applied model selection criteria in order to select the most favored parametrization of $q(z)$ among those that they studied in their paper \cite{ElgaroyMultamaki2006}. This is definitely better than just choosing one parametrization in a fully blind way, but nevertheless we deem that this does not completely solve the problem, since there can be several parametrizations leading to different associated values of e.g. $q_0$ that in terms of model selection criteria offer a similar efficiency. Picking just one form of $q(z)$ might therefore lead us still to biased conclusions and to underestimate the uncertainties of the measured quantities, even if we use model-selection criteria to carry out our choice. Thus, depending on the physical question we are interested to answer we are still forced to search for an alternative approach which does not depend on particular parametrizations of $q(z)$ or any other alternative cosmographical function. In the next section we describe the WFR method, a generalization of the procedure applied in \cite{ElgaroyMultamaki2006} which is able to mitigate even more the problem, and to go one step further concerning the model-independence of these kind of analyses.

%%%%%%%%%%%%%%%%%%%%%%%%%%%%%%%%%%%%%%%%%%%%%%%%%%%%%%%%%%%%%% 
%%%%%%%%%%%%%%%%%%%%%%%%%%%%%%%%%%%%%%%%%%%%%%%%%%%%%%%%%%%%%% 
%%%%%%%%%%%%%%%%%%%%%%%%%%%%%%%%%%%%%%%%%%%%%%%%%%%%%%%%%%%%%% 

\subsection{Reconstruction of $E(z)$ with the WFR method}\label{sec:WFRdes}

Those works in which the authors choose a particular truncated cosmographical series to carry out the fitting analysis are also susceptible to the problems that we have exposed in the preceding subsection. In this case the situation is not very different from choosing a concrete parametrization of $q(z)$. To understand why, let us focus on the cosmographical expansions (\ref{eq:EexpZ}) and (\ref{eq:EexpY}). If we cut these series at a given order and apply (\ref{eq:q(z)1}) we can obtain the form of $q(z)$ associated to the aforesaid truncated series of $E(z)$. The problem we encounter is therefore completely analogous to the one described in the last part of Sect. \ref{sec:Motivation}. Now we will try to alleviate it in the cosmographical context of (\ref{eq:EexpZ}) and (\ref{eq:EexpY}) and the WFR method. The mathematical structure of these expansions of the Hubble rate have something in common: they are built linear in the coefficients $c_i$,
\begin{equation}
E(z)=1+\sum_{i=1}^{\infty}c_ig_i(z)\,,
\end{equation}
where the $c_i$'s are constants that can be expressed in terms of the cosmographical parameters, i.e. $q_0$, $j_0$, $s_0$, etc.\footnote{For the sake of clarity, we remark that we will refer to the $c_i$'s as coefficients of the expansion, and to the $q_0$, $j_0$, $s_0$, etc. as the cosmographical parameters.}, and the $g_i(z)$'s are functions of the redshift with a very simple structure, $g_i(z)=[g_1(z)]^i$, being $g_1(z)=z$ in (\ref{eq:EexpZ}) and $g_1(z)=z/(1+z)$ in (\ref{eq:EexpY}). They are usually referred to as {\it basis} functions. Instead of relying on one particular truncated series, 
\begin{equation}\label{eq:genericE}
E_J(z)=1+\sum_{i=1}^{J}c_ig_i(z)\,,
\end{equation}
and thus set $J$ to a concrete value in our study, we opt to incorporate the information about all the nested expansions (obtained by changing $J$) in order to skip the problem of choosing just one among them in the fitting analysis. Let us call $M_1$, $M_2$,..., $M_{N}$ the cosmographical expansions of order $J=1$, $2$,..., $N$, respectively, with $N$ being the number of data points used in the analysis. That is, let us conceive each expansion as a different model, and compute the probability density associated to the fact of having a certain shape for the Hubble rate as follows,
\begin{equation}
P[E(z)]=k\cdot[P(E(z)|M_1)P(M_1)+...+P(E(z)|M_{N})P(M_{N})]\,,
\end{equation}
where $k$ is just a normalization constant that must be fixed by imposing
\begin{equation}
\int[\mathcal{D}E]\,P[E(z)]=1\,.
\end{equation}
Taking into account that 
\begin{equation}
\int[\mathcal{D}E]\,P(E(z)|M_J)=1\quad \forall J\in[1,N]
\end{equation}
and  
\begin{equation}\label{eq:normRel}
\sum_{J=1}^{N} P(M_J)=1\,,
\end{equation}
we find $k=1$ and therefore:
\begin{equation}
P[E(z)]=\sum_{J=1}^{N}P(E(z)|M_J)P(M_J)\,.
\end{equation}
We now denote $M_*$ as the most probable model and rewrite the last expression as follows,
\begin{equation}\label{eq:penExp}
P[E(z)]=P(M_*)\sum_{J=1}^{N}P(E(z)|M_J)\frac{P(M_J)}{P(M_*)}\,,
\end{equation}
where $\frac{P(M_J)}{P(M_*)}$ can be identified with the Bayes ratio $B_{J*}$, i.e. the ratio of evidences
\begin{equation}\label{eq:BayesRatio}
B_{J*}=\frac{\mathcal{E}_J}{\mathcal{E}_*}=\frac{\int\mathcal{L}(\mathcal{D}|\vec{c}_J)\pi(\vec{c}_J)d\vec{c}_J}{\int\mathcal{L}(\mathcal{D}|\vec{c}_*)\pi(\vec{c}_*)d\vec{c}_*}\,,
\end{equation}
with $\mathcal{L}(\mathcal{D}|\vec{c}_J)$ being the likelihood, which is a function of the coefficients entering the model $J$, $\vec{c}_J$, and the data set $\mathcal{D}$ (which of course is common for all the models), and $\pi(\vec{c}_J)$ being the prior for the coefficients, see e.g. \cite{DEbook,AmendolaNotes}. $M_*$ is formally defined as the model with largest evidence in the whole set $\{M_J\}$. Using (\ref{eq:normRel}) one finds
\begin{equation}
P(M_*)=\left(\sum_{J=1}^{N}B_{J*}\right)^{-1}\,,
\end{equation}
so (\ref{eq:penExp}) can be finally written as
\begin{equation}\label{eq:finExp}
P[E(z)]=\frac{\sum\limits_{J=1}^{N}P(E(z)|M_J)B_{J*}}{\sum\limits_{J=1}^{N}B_{J*}}\,.
\end{equation}
This is the central expression of the weighted function regression method, where the weights are directly given by the Bayes factors. Notice that making use of (\ref{eq:finExp}) we can compute the (weighted) moments and related quantities too. For instance, the weighted mean and variance read,
\begin{equation}\label{eq:MeanTotal}
\bar{E}(z)=\int [\mathcal{D}E]P[E(z)]E(z)=\frac{\sum\limits_{J=1}^{N}\bar{E}_{J}(z)B_{J*}}{\sum\limits_{J=1}^{N}B_{J*}}\,,
\end{equation}
\begin{equation}\label{eq:VarianceTotal}
\sigma^2(z)=\int [\mathcal{D}E]P[E(z)](E(z)-\bar{E}(z))^2=\frac{\sum\limits_{J=1}^{N}[\sigma_J^2(z)+(\bar{E}_{J}(z))^2]B_{J*}}{\sum\limits_{J=1}^{N}B_{J*}}-(\bar{E}(z))^2\,,
\end{equation}
where $\bar{E}_{J}(z)$ and $\sigma_J(z)$ are the mean and standard deviation computed in the model $J$ (we will show in Sect. \ref{sec:evidences} how to calculate them analytically in the case under study). We remark here that these functions will differ in general from the best-fit function and its associated $68.3\%$ c.l. bands due to the possible deviations from Gaussianity encountered in the final reconstructions. It is also possible to estimate the effective number of parameters in the final reconstruction, using
\begin{equation}
N_{\rm eff}=\frac{\sum\limits_{J=1}^{N}JB_{J*}}{\sum\limits_{J=1}^{N}B_{J*}}\,.
\end{equation}
The  machinery explained in this subsection was already employed in \cite{GomezAmendola2018} to reconstruct the Hubble function in the light of the CCH and Pantheon+MCT SnIa data, using (\ref{eq:MeanTotal}) and (\ref{eq:VarianceTotal}), and evaluating the Bayes ratio approximately with the help of the Akaike \cite{Akaike} and Bayesian \cite{Schwarz} information criteria as explained in Sect. 4.2 of our past paper. Now we aim to reconstruct $E(z)$, $q(z)$ and $j(z)$ using the weighted function regression formalism too, but improving the methodology in two important aspects with respect to \cite{GomezAmendola2018}, namely: (i) here we will compute not the mean and variance of these functions, but the best-fit and corresponding exact $1\sigma$ confidence regions; and (ii) we will calculate the exact Bayes ratios with the formula (\ref{eq:BayesRatio}), instead of using approximations of it. In the case under study it is possible to compute the exact expressions for the evidences analytically because the fitting functions (\ref{eq:genericE}) are in all cases, i.e. $\forall{J}$, linear in the parameters $c_i$ and, in addition, the data on $E(z)$ described in Sect. \ref{sec:Data} are Gaussian-distributed in very good approximation. In the next subsection we explicitly derive the formula for the evidence, which plays a very important role in the WFR method, since it is in charge of controlling the weight of the various models in the final distribution (\ref{eq:finExp}). 

%%%%%%%%%%%%%%%%%%%%%%%%%%%%%%%%%%%%%%%%%%%%%%%%%%%%%%%%%%%%%% 
%%%%%%%%%%%%%%%%%%%%%%%%%%%%%%%%%%%%%%%%%%%%%%%%%%%%%%%%%%%%%% 
%%%%%%%%%%%%%%%%%%%%%%%%%%%%%%%%%%%%%%%%%%%%%%%%%%%%%%%%%%%%%% 

\subsection{Computation of evidences and other quantities of interest}\label{sec:evidences}

We begin this subsection reviewing the main expressions needed for fitting Gaussian-distributed data with functions that are linear in the coefficients, as in the case that concerns us. If we have a collection $\mathcal{D}=\{(z_\mu,y_\mu),\,\mu=1,...,N,\,N\geq J\}$ of Gaussian-distributed data points with covariance matrix $C$, and we want to fit (\ref{eq:genericE}) to them we have to maximize the likelihood
\begin{equation}\label{eq:likelihood}
\mathcal{L}(\mathcal{D}|\vec{c})=\frac{1}{(2\pi)^{N/2}\sqrt{|C|}}e^{-\frac{1}{2}[y_\mu-E(z_\mu;\vec{c})]C^{-1}_{\mu\beta}[y_\beta-E(z_\beta;\vec{c})]}
\end{equation}
with respect to the elements of the vector of coefficients $\vec{c}$. We have omitted here the subscripts $J$ for simplicity, but it is important to keep in mind we are referring to a particular model $M_J$, so the theoretical expression for the Hubble rate is characteristic of this concrete model, and so are the covariance matrices and mean values of the coefficients that will appear in the subsequent formulas for both, the likelihood and the prior distributions. Notice also that in the last formula we are using the Einstein summation convention, as we will do in all the forthcoming expressions unless stated otherwise. We use Greek letters for indexes labeling data points, and Latin ones for those labeling the terms of $E(z)$, as in (\ref{eq:genericE}). Due to the linearity of the latter in the coefficients it is possible to rewrite the likelihood (\ref{eq:likelihood}) as a multivariate Gaussian distribution for the coefficients too, i.e.
\begin{equation}\label{eq:likelihood2}
\mathcal{L}(\mathcal{D}|\vec{c})=\frac{e^{-\chi^2_{\rm min}/2}}{(2\pi)^{N/2}\sqrt{|C|}} e^{-\frac{1}{2}(c_i-\bar{l}_i)F_{ij}(c_j-\bar{l}_j)}\,,
\end{equation}
where $\chi^2_{\rm min}=\chi^2(\vec{c}=\vec{\bar{l}})$,
\begin{equation}\label{eq:covParam}
F_{ij}=G_\mu^i C_{\mu\beta}^{-1} G_\beta^j
\end{equation}
is the inverse covariance matrix of the coefficients, also known as Fisher matrix, $G_\mu^i\equiv g_i(z_\mu)$, and
\begin{equation}\label{eq:meanParam}
\bar{l}_i=y_\mu C_{\mu\beta}^{-1}G_\beta^jF^{-1}_{ij}
\end{equation}
is the mean value derived from the likelihood for the coefficient $c_i$. Equipped with these tools, it is straightforward to compute the mean function $\bar{E}_J(z)$ and covariance matrix ${\rm cov}[E_J(z),E_J(z^\prime)]$ in a given model $M_J$. It can be done as follows (here we write again the sum symbols explicitly),
\begin{equation}\label{eq:meanRecFunc}
\bar{E}_J(z)=1+\sum_{i=1}^{J}\bar{l}_ig_i(z)\,,
\end{equation}
\begin{equation}\label{eq:covRecFunc}
{\rm cov}[E_J(z),E_J(z^\prime)]=\sum_{i,j=1}^{J} F^{-1}_{ij}g_i(z)g_j(z^\prime)\,.
\end{equation}
The variance of the reconstructed function in model $J$ is just $\sigma_J^2(z)={\rm cov}[E_J(z),E_J(z)]$. These expressions are involved in the computation of (\ref{eq:MeanTotal}) and (\ref{eq:VarianceTotal}). Now, we have all the ingredients to derive the compact formula for the posterior distribution in one particular model $M_J$. It can also be found in many other references, as in \cite{Nesseris2013,DEbook,AmendolaNotes}, but we add this information here too for completeness. The product of the Gaussian prior,  
\begin{equation}\label{eq:prior}
\pi(\vec{c})=\frac{1}{(2\pi)^{J/2}\sqrt{|P|}} e^{-\frac{1}{2}(c_i-\bar{p}_i)P^{-1}_{ij}(c_j-\bar{p}_j)}\,,
\end{equation}
with the likelihood (\ref{eq:likelihood2}) can be written as follows,
\begin{equation}\label{eq:product}
\mathcal{L}(\mathcal{D}|\vec{c})\pi(\vec{c})=\frac{e^{-\frac{1}{2}(\chi^2_{\rm min}+\bar{l}_i\bar{l}_jF_{ij}+\bar{p}_i\bar{p}_jP^{-1}_{ij}-\bar{d}_i\bar{d}_jD^{-1}_{ij})}}{(2\pi)^{(J+N)/2}\sqrt{|P||C|}}e^{-\frac{1}{2}(c_i-\bar{d}_i)D^{-1}_{ij}(c_j-\bar{d}_j)}\,,
\end{equation}
with 
\begin{equation}\label{eq:PostCovMatrix}
D^{-1}_{ij}=F_{ij}+P^{-1}_{ij}
\end{equation}
\noindent being the inverse of the posterior covariance matrix, and  
\begin{equation}\label{eq:PostMean}
\bar{d}_k=D_{ki}(F_{ij}\bar{l}_j+P^{-1}_{ij}\bar{p}_{j})
\end{equation}
the posterior mean of the coefficient $c_k$. The latter coincides with the best-fit value, since the posterior distribution is also a multivariate Gaussian. The integration of (\ref{eq:product}) with respect to the coefficients of the model, $\vec{c}$, is straightforward and leads us to the final expression for the evidence that enters the formula of the Bayes ratio (\ref{eq:BayesRatio}),
\begin{equation}\label{eq:evidence}
\mathcal{E}=\frac{1}{(2\pi)^{N/2}\sqrt{|C|}}\sqrt{\frac{|D|}{|P|}}e^{-\frac{1}{2}(\chi^2_{\rm min}+\bar{l}_i\bar{l}_jF_{ij}+\bar{p}_i\bar{p}_jP^{-1}_{ij}-\bar{d}_i\bar{d}_jD^{-1}_{ij})}\,.
\end{equation}
As it is explained in e.g. \cite{DEbook,AmendolaNotes}, the evidence depends on three factors: (i) the ability of the model to fit the data of the likelihood (\ref{eq:likelihood}). This fixes the value of $\chi^2_{\rm min}$; (ii) the relative difference between the constraints set by the likelihood and the prior. This fixes the ratio $|D|/|P|$; and (iii) the distance in parameter space between the best-fit values preferred by the likelihood and those that are preferred by the prior, which affects the computation of $\bar{l}_i\bar{l}_jF_{ij}+\bar{p}_i\bar{p}_jP^{-1}_{ij}-\bar{d}_i\bar{d}_jD^{-1}_{ij}$. The evidence automatically integrates Occam's razor criterion in its definition, since the addition of extra parameters (when we take a model with $J^\prime>J$) reduces the value of $\mathcal{E}$ when they are constrained by the likelihood in a comparable way (or better) than by the prior. This means that the weights in our WFR method are built through (\ref{eq:BayesRatio}) according to this criterion as well. On the one hand, adding more parameters reduces the value of $\chi^2_{\rm min}$ (if these parameters are effective enough, of course) and this increases the value of $\mathcal{E}$; on the other, there is the corresponding penalization, as mentioned before, so there exists a competition between these two opposite effects.

%%%%%%%%%%%%%%%%%%%%%%%%%%%%%%%%%%%%%%%%%%%%%%%%%%%%%%%%%%%%%%

\begin{figure}[t!]
\begin{center}
\includegraphics[scale=0.5]{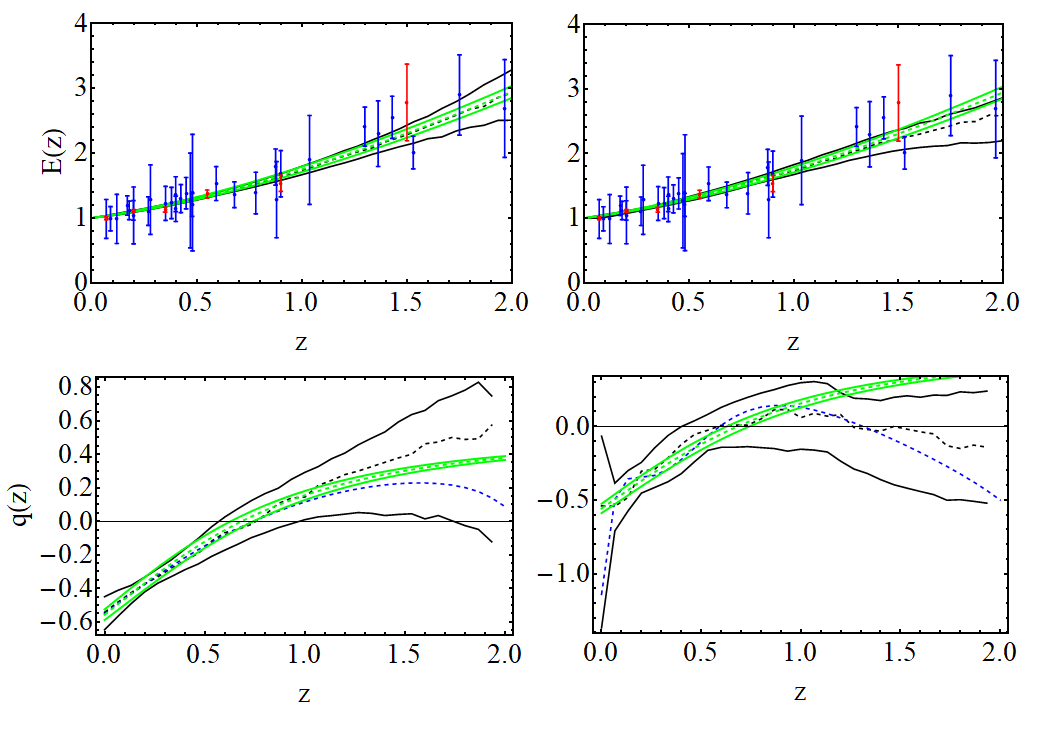}
\caption{In black, best-fit values and $1\sigma$ uncertainties of the reconstructed functions $E(z)$ (top row) and the derived $q(z)$ (bottom row) that are obtained using the WFR method as described in Sects. \ref{sec:WFRdes} and \ref{sec:evidences}, based on the expansions (\ref{eq:EexpZ}) (plots in the left column) and (\ref{eq:EexpY}) (in the right one), and making use of the Pantheon+MCT (prior) and CCH (likelihood) data. These data points are incorporated to the upper plots with red and blue error bars, respectively. The blue-dashed curves in the lower plots refer to the mean of $q(z)$. We also show (in green) the central curves and $1\sigma$-bands for the $\Lambda$CDM model, obtained using the best-fit values of Table 3. See the related comments in the main text.}
\end{center}
\end{figure}

%%%%%%%%%%%%%%%%%%%%%%%%%%%%%%%%%%%%%%%%%%%%%%%%%%%%%%%%%%%%%%

Formulas (\ref{eq:PostCovMatrix}) and (\ref{eq:PostMean}) are fully symmetric under the interchange of the prior and likelihood covariance matrices and best-fit values. Therefore, once we divide the data into two parts, the posterior best-fit values and associated uncertainties for a given model do not depend at all on which of these parts is used to build the prior and which is used to construct the likelihood. There is a kind of freedom at this point. It is important to remark, though, that the prior distribution cannot be built with data that already take part of the likelihood, since this would produce an unwanted double counting which would make the analysis inconsistent. For example, we are not allowed to use the data from the JLA SnIa compilation as a prior and the Pantheon ones for the likelihood, just because the latter contains the former (cf. Sect. \ref{sec:2p1}), and we would obtain biased constraints on the parameters of the model under study. Conversely, the evidence (\ref{eq:evidence}) is not symmetric under the aforementioned interchange, so the weights used in the WFR method are sensitive to it. We will analyze its impact in Sect. \ref{sec:results}. 

For any model $M_J$, we can use a multivariate Gaussian sampler, using e.g. {\it Mathematica} \cite{Mathematica}, in order to produce a list of vectors $\vec{c}$ following the posterior distribution (\ref{eq:product}) with mean vector (\ref{eq:PostMean}) and inverse covariance matrix (\ref{eq:PostCovMatrix}). For each of these vectors we can compute the functions of interest, $E_J(z)$, $q_J(z)$, and $j_J(z)$. Notice that we do not need to carry out any Monte Carlo algorithm to do that. Thanks to the fact of having the analytical expressions (\ref{eq:PostCovMatrix}) and (\ref{eq:PostMean}) we can generate histograms of the aforementioned functions at the wanted redshifts without loosing the computational time that we would have to expend with a Monte Carlo exploration of the parameter space. In this way we can obtain the same level of statistics roughly three times faster, because we do not have to throw away the $\sim 60\%-70\%$ of the points, as it is done in a typical Markov chain Monte Carlo run. We remark that the analytical obtention of (\ref{eq:PostCovMatrix}) and (\ref{eq:PostMean}) has been possible because the data is Gaussian-distributed (cf. Sect. \ref{sec:Data}) and also because the fitting functions (\ref{eq:genericE}) are linear in the coefficients $c_i$. The analytical computation of the evidence with formula (\ref{eq:evidence}) also allows us to save valuable computational time, since we can avoid the calculation of the corresponding integral in the $J$-dimensional parameter space, cf. the formula (\ref{eq:BayesRatio}). Once we obtain the histograms with the values of the various functions evaluated at several redshifts for all the models, we can construct the corresponding join distribution (as a histogram, of course) for each function and redshift using the weights computed before for each model $M_J$ and following (\ref{eq:finExp}), to finally derive the WFR-reconstructed shapes of the quantities of interest. We present our results in the next section.

%%%%%%%%%%%%%%%%%%%%%%%%%%%%%%%%%%%%%%%%%%%%%%%%%%%%%%%%%%%%%%

\begin{figure}[t!]
\begin{center}
\includegraphics[scale=0.5]{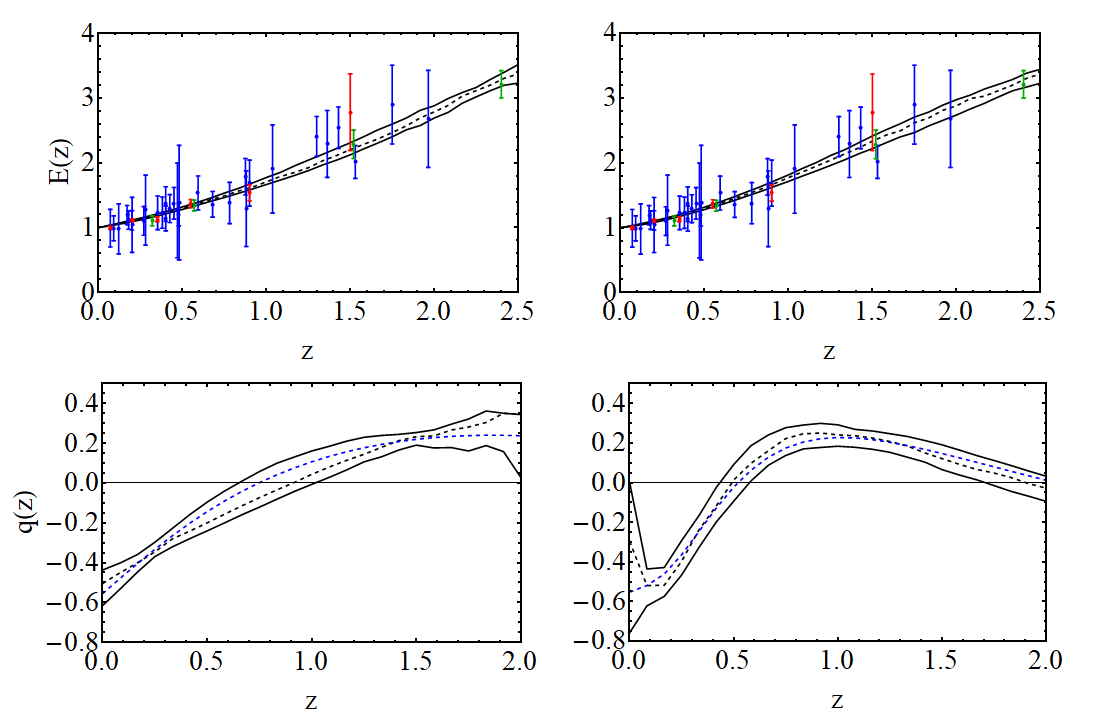}
\caption{As in Fig. 3, but incorporating the BAOs data to the likelihood. They are depicted in green in the upper plots.}
\end{center}
\end{figure}

%%%%%%%%%%%%%%%%%%%%%%%%%%%%%%%%%%%%%%%%%%%%%%%%%%%%%%%%%%%%%% 

%%%%%%%%%%%%%%%%%%%%%%%%%%%%%%%%%%%%%%%%%%%%%%%%%%%%%%%%%%%%%% 
%%%%%%%%%%%%%%%%%%%%%%%%%%%%%%%%%%%%%%%%%%%%%%%%%%%%%%%%%%%%%% 
%%%%%%%%%%%%%%%%%%%%%%%%%%%%%%%%%%%%%%%%%%%%%%%%%%%%%%%%%%%%%%

\section{Results and discussion}\label{sec:results}

We start analyzing and discussing the results that we obtain applying the methodology explained in the previous section to the direct reconstruction of the Hubble rate and the derived deceleration parameter. In Fig. 3 we show the results for the case in which we use the Pantheon+MCT and CCH data, being the former employed to build the prior, and the latter to construct the likelihood (later on we will analyze the impact of this choice in detail). In the left column of this figure we present the results obtained with the WFR method when use is made of the expansion in the redshift (\ref{eq:EexpZ}), whereas those of the right column are obtained using (\ref{eq:EexpY}), which is built in the $y$-variable. It is obvious that the reconstructed Hubble rates are in both cases very similar, not only concerning the central values, but also the $1\sigma$-bands. In contrast, although the reconstructed shapes for $q(z)$ are completely compatible, one can appreciate important differences concerning the size of the error bands in some regions of the covered redshift range. This is palpable around $z=0$. Using (\ref{eq:EexpZ}) we obtain $q_0=-0.55^{+0.09}_{-0.11}$, whereas using (\ref{eq:EexpY}) $q_0=-0.54^{+0.47}_{-0.83}$ (both at $1\sigma$ c.l.), so the latter is compatible with $0$ at only $\sim 1\sigma$. This is clearly pointing out some kind of defect affecting (at least) one of the two expansions. We are using the WFR method precisely to remove the model and parametrization-dependence that is inherent to many other analyses in the literature, but concerning the shape of $q(z)$ we see that this parametrization-dependence still persists, we have not been able to get rid of it. However, we can already suspect from the discussion held in Sect. \ref{sec:cosmography} that the problematic reconstruction is the one derived from (\ref{eq:EexpY}). Notice moreover that the $\Lambda$CDM curves (which are drawn in green in all the plots of Fig. 3) are in all cases fully compatible at $<1\sigma$ with the reconstructed functions. They are actually contained inside the reconstructed bands. The only exception is at a small region at $z>1.2$ of the lower right plot, which is compatible not at one, but at two sigmas. The $1\sigma$-bands, though, are much smaller in the $\Lambda$CDM than in the WFR-reconstructions. This is the price we have to pay for the model-independence of our analysis (which, as already said, seems not to be yet parametrizarion-independent, see the subsequent comments below).  

In Fig. 4 we show the results that we obtain by also considering the BAOs information. The problem mentioned above does not disappear neither in this case. The reconstructed functions $E(z)$ are now even more resonant than before, and the error bands decrease for the Hubble rate and $q(z)$, as expected, but the reconstructed deceleration parameter still suffers from the same problems mentioned before. Now, $q_0=-0.51^{+0.08}_{-0.10}$ with (\ref{eq:EexpZ}), and with (\ref{eq:EexpY}) it is still compatible with $0$ at $1\sigma$. The reasons of such discrepancy are not difficult to understand from a mathematical point of view. The first one has to do with the discussion of Sect. \ref{sec:cosmography}. The second is intimately related to the former. With the truncated low-order expansions derived from (\ref{eq:EexpY}) the cosmographical parameters must depart significantly from the true values in order to fit correctly the data points at the largest redshifts. Moreover, these data points have a much greater impact on $q_0$ than the one encountered in the expansions obtained from (\ref{eq:EexpZ}), as it will be shown now. In order to ease the explanation we restrict ourselves to the case in which we truncate the series (\ref{eq:EexpZ}) and (\ref{eq:EexpY}) at second order, yielding
\begin{equation}\label{eq:expansions}
E_z(z)-1=az+bz^2\quad ;\quad E_y(z)-1=\tilde{a}\frac{z}{1+z}+\tilde{b}\left(\frac{z}{1+z}\right)^2\,,
\end{equation}
%

%%%%%%%%%%%%%%%%%%%%%%%%%%%%%%%%%%%%%%%%%%%%%%%%%%%%%%%%%%%%%%
\begin{table}
\begin{center}
\begin{scriptsize}
\begin{tabular}{|c|c|c|c|c|c|}

\multicolumn{1}{c}{$J$} & \multicolumn{1}{c}{$q_{0,J}$} & \multicolumn{1}{c}{$z_{t,J}$} & \multicolumn{1}{c}{$j_{0,J}$} & \multicolumn{1}{c}{$\mathcal{E}_J$} & \multicolumn{1}{c}{${\rm w}_i\equiv\frac{\mathcal{E}_J}{\sum_i\mathcal{E}_i}$ }
\\\hline
$2$ & $-0.50^{+0.05}_{-0.04}$ & $0.91^{+0.09}_{-0.06}$ & $0.59^{+0.13}_{-0.12}$ & 318.57 & 0.52 \\ \hline
$3$ & $-0.64^{+0.09}_{-0.07}$ & $0.58^{+0.17}_{-0.07}$ & $1.48^{+0.25}_{-0.51}$ & 274.48 & 0.45 \\  \hline
$4$ & $-0.62^{+0.12}_{-0.17}$ & $0.60^{+0.15}_{-0.13}$ & $1.5^{+1.3}_{-1.0}$ & 19.74 & 0.03 \\\hline
\end{tabular}
\end{scriptsize}
\caption{Current values of the deceleration and jerk parameters, transition redshift, evidence and corresponding weight for each model involved in the same WFR-reconstructions of the left plots in Fig. 4, those that make use of (\ref{eq:EexpZ}). The model containing only the linear term in $z$, i.e. with $J=1$, and all the models with $J>4$ do not contribute significantly to the final reconstruction because they have a negligible weight in the global distribution (\ref{eq:finExp}). We have not included their information in this table. The effective number of degrees of freedom is $N_{\rm eff}=2.52$, and the final reconstruction leads to: $q_0=-0.51^{+0.08}_{-0.10}$, $z_t=0.90^{+0.12}_{-0.25}$, and $j_0=0.59^{+0.64}_{-0.12}$.}
\end{center}
 \label{tab:info}
\end{table}
%%%%%%%%%%%%%%%%%%%%%%%%%%%%%%%%%%%%%%%%%%%%%%%%%%%%%%%%%%%%%%  

\noindent respectively. The point is the following. The data at $z>1$ have a greater impact on the coefficient $\tilde{a}$ rather than on $a$, just because the relative weight of the squared term in $E_y(z)$ (the one containing $\tilde{b}$) is much lower than the one in $E_z(z)$ and, therefore, the influence of those data points at higher redshifts on the first terms of the right-hand side of the expansions (\ref{eq:expansions}) is much bigger in $E_y(z)$ than in $E_z(z)$. For instance, at $z=2$ we have $E_z(2)-1=2(a+2b)$ and $E_y(2)=2(3\tilde{a}+2\tilde{b})/9$, so the relative weight of $\tilde{a}$ is three times larger than the one of $a$ at this redshift. By comparing (\ref{eq:EexpZ}) and (\ref{eq:EexpY}) with (\ref{eq:expansions}) one can see that the relation between $q_0$ and the $a$'s are exactly the same, i.e. $q_0=a-1$ and $q_0=\tilde{a}-1$, respectively, so $q_0$ will depend more strongly on the data points at high redshifts in the parametrization (\ref{eq:EexpY}) than in (\ref{eq:EexpZ}). In fact, notice that if we had data points at very high redshifts, let us say at $z\to \infty$, then these data would not influence $a$ at all, whereas $\tilde{a}$ would have the same weight as $\tilde{b}$. This situation is clearly anomalous and is probably telling us that the parametrization (\ref{eq:EexpY}) is quite unpractical and unable to grasp properly the correct physical behavior of the underlying function, i.e. the function we aim to reconstruct, just because $q_0$ is too sensitive to the data at very high redshifts. We must be more confident on the constraints for $q_0$ obtained through (\ref{eq:EexpZ}) than the ones obtained through (\ref{eq:EexpY}). Actually, it is quite abnormal not to find any evidence for the current accelerated phase of the Universe with (\ref{eq:EexpY}). Conversely, with the parametrization (\ref{eq:EexpZ}) we find very strong evidence applying a full Bayesian approach. We can repeat the same procedure used to reconstruct $E(z)$, but fixing $q_0=0$. We compute then the sums of the evidences derived from all the models $E_J(z)$ when $q_0=0$, i.e. $\mathcal{E}_J(q_0=0)$, and when $q_0\ne 0$ and left free in the fitting analyses, i.e. $\mathcal{E}_J(q_0\ne 0)$. Then we calculate the Bayes ratio using these sums, as follows,
%%%%%%%%%%%%%%%%%%%%%%%%%%%%%%%%%%%%%%%%%%%%%%%%%%%%%%%%%%%%%%

\begin{figure}
\begin{center}
\includegraphics[scale=0.7]{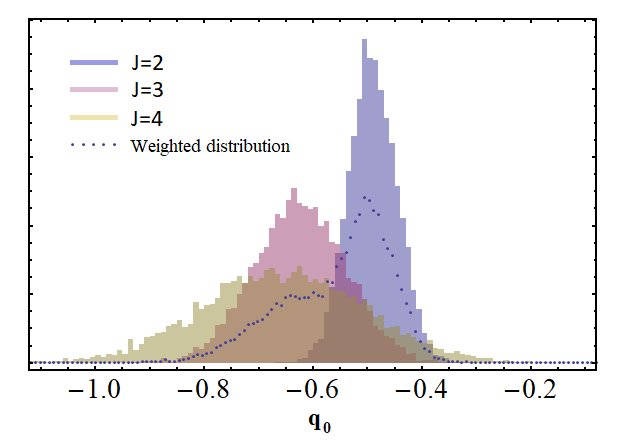}
\caption{Individual distributions for $q_0$ for the various models that contribute non-negligibly to the final WFR-reconstructions of Fig. 4 (plots on the left), cf. also Table 5. We also plot the weighted distribution, which is built from the individual ones analogously to (\ref{eq:finExp}).}
\end{center}
\end{figure}

%%%%%%%%%%%%%%%%%%%%%%%%%%%%%%%%%%%%%%%%%%%%%%%%%%%%%%%%%%%%%%

%
\begin{equation}\label{eq:BR}
B=\frac{\sum\limits_{J}\mathcal{E}_J(q_0\ne 0)}{\sum\limits_{J}\mathcal{E}_J(q_0=0)}=1137\rightarrow \ln B=7.04\,.
\end{equation}
According to Jeffreys' scale (see e.g. Refs. \cite{KassRaftery1995,DEbook,AmendolaNotes,Jeffreys}), this is pointing towards a very strong evidence in favor of an accelerated Universe \footnote{Using the value of $H_0$ provided in \cite{RiessH02018} -- instead of $H_0=(70\pm 5)$ ${\rm km\,s^{-1}Mpc^{-1}}$ -- in the prior employed to convert the original CCH+BAOs data into data on the Hubble rate (see Sect. \ref{sec:2p2}), we find $q_0=-0.52^{+0.05}_{-0.09}$ and $\ln B=7.96$. Using the one derived from the fitting analysis of the $\Lambda$CDM carried out by the Planck Collaboration (2018) \cite{Planck2018} with the TT+lowE CMB data we obtain $q_0=-0.46^{+0.06}_{-0.15}$ and $\ln B=6.07$. In both cases the strong level of evidence is maintained, so our conclusion does not depend on this. The same happens if we use the prior on $r_s(z_d)$ from \cite{Heavens2014} instead of the one from \cite{Verde2017}. In this case one gets $q_0=-0.48^{+0.06}_{-0.09}$ and, again, very strong evidence in favor of the current speed-up of the Universe.}, since $\ln B>5$. Here the BAOs data play a crucial role to enhance the confidence level of our result. We have checked that if we remove them from our data set $\ln B=2.83$ and, hence, the evidence decreases up to a moderate level, just because in this case $1<\ln B<3$. This in stark contrast with the results found e.g. in the context of the $\Lambda$CDM (cf. Table 3), in which one finds $q_0=-0.554\pm0.032$ using only the SnIa and the CCH data. This $17\sigma$-evidence for the positive acceleration of the Universe is very far away from the more conservative one that we have inferred from our more model-independent approach, which is roughly three times smaller.

%%%%%%%%%%%%%%%%%%%%%%%%%%%%%%%%%%%%%%%%%%%%%%%%%%%%%%%%%%%%%%

\begin{figure}
\begin{center}
\includegraphics[scale=0.55]{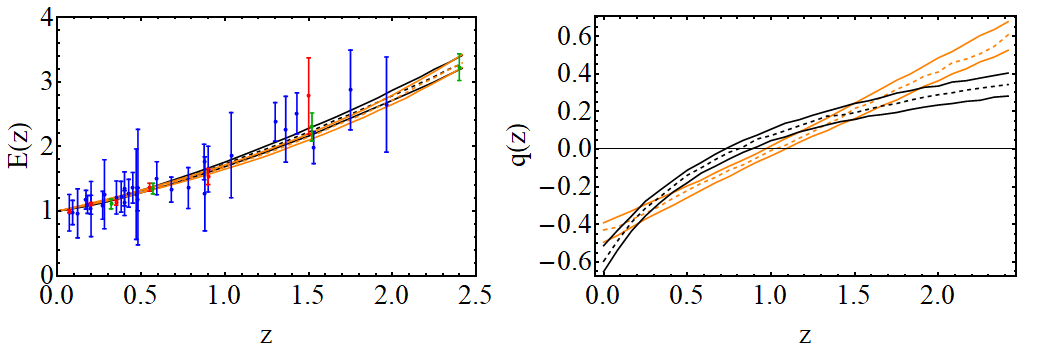}
\caption{{\it Left plot:} Reconstructed Hubble rates and $1\sigma$-bands obtained with the WFR method, using the CCH+BAO (prior) and Pantheon+MCT (likelihood) data, and the parametrizations of $q(z)$ (\ref{eq:ExpqZ}, in orange) and (\ref{eq:ExpqY}, in black). The data points are depicted with the same colors of Fig. 4; {\it Right plot:} The corresponding reconstructed deceleration parameters.}
\end{center}
\end{figure}

%%%%%%%%%%%%%%%%%%%%%%%%%%%%%%%%%%%%%%%%%%%%%%%%%%%%%%%%%%%%%%

In Table 5 we provide some relevant information about the individual models (\ref{eq:EexpZ}) employed in the WFR-reconstructions of $E(z)$ and $q(z)$ that we have plotted in the left column of Fig. 4. Actually, only three of these models play an important role in these reconstructions: the second, third and fourth-order polynomials of $E(z)$. The other ones are strongly suppressed either because they are completely unable to fit correctly the data, as the linear expansion of $E(z)$, with $J=1$, or because they receive a very important penalization for using too many parameters. It is the case of those models with $J>4$. Remarkably, the best-fit values of the parameters $q_0$, $j_0$, and also $z_t$, obtained with the global WFR-distribution (\ref{eq:finExp}) are very close to those obtained in the model with highest evidence, i.e. the one with $J=2$. Although the addition of the other models (basically those with $J=3,4$) in the weighted sum only shifts very slightly the central values of the parameters, they increase the total uncertainty (compare the values in Table 5 with those provided in its caption). 

In regards to the central values of $j_0$ listed in Table 5 for the various models (i.e. for the various $J$s), the reader will observe that they are quite different. It is very important to notice, though, that in terms of the number of sigmas the statistical differences are not too significant. For instance, the value obtained with the model $J=4$, $j_0=1.5^{+1.3}_{-1.0}$, is fully compatible at $<1\sigma$ c.l. with the values obtained with $J=2$ and $J=3$, which read respectively, $j_0=0.59^{+0.13}_{-0.12}$ and $j_0=1.48^{+0.25}_{-0.51}$. Something similar is found when one compares the last two values. If one takes the upper uncertainty of the former and the lower uncertainty of the latter one founds a difference between the best-fit values of $1.7\sigma$, which is not very significant from a statistical point of view. This discussion has to do with the root of the problem we are trying to cope with in this paper, which has been described in detail in Sect. \ref{sec:Motivation} through several examples in the context of specific cosmological models and parametrizations of the deceleration parameter. Choosing different truncation orders of the cosmographical series employed to fit the data leads to quite different estimations of the kinematic quantities. The differences affect in general not only the central values, but also the size of the error bars. The WFR method allows to mitigate this problem by weighting in a consistent way the contribution of the various truncated expansions to the final reconstructed shape, as it is described in detail in Sects. \ref{sec:WFRdes} and \ref{sec:evidences}. The value of $j_0$ that results from the reconstructed function $j(z)$, $j_0=0.59^{+0.64}_{-0.12}$ (cf. the caption of Table 5, and Fig. 8), is not in tension with the value for any model. Something similar happens with the $q_0$-values for models $J=2$ and $J=3$. They are in very mild tension, of only $1.4\sigma$. The weighted value, though, is completely compatible with them, as expected.

In order to better visualize the interplay of the various models in the generation of the final output we have also included Fig. 5. There we show the Gaussian shape of the individual distributions for $q_0$ obtained in the models with $J=2,3,4$, together with the weighted distribution built following (\ref{eq:finExp}). The latter is highly non-Gaussian. 

We have also studied what is the impact of the systematic errors that we have introduced in Sect. \ref{sec:Data} to account for the choice of SPS model in the CCH data. More concretely, we have checked that considering a more conservative systematic uncertainty given not by the difference of the central values for $H^{\rm ori}(z_i)$ (cf. formula \eqref{eq:Cor2}) but by half the difference of these quantities we obtain results which are very similar to the ones reported in Figs. 4-5 and Table 5. Using the same election of data to construct the prior and the likelihood we obtain $q_0=-0.49^{+0.07}_{-0.09}$, $z_t=0.89^{+0.11}_{-0.16}$, and $j_0=0.67^{+0.42}_{-0.13}$. Comparing these results with those reported in the caption of Fig. 4 one can see that the only effect is a little shift of the central values (which is negligible in front of the uncertainties' size) and a decrease of the error bars. The latter is very small for the deceleration parameter, whereas is bigger for the lower uncertainty of the transition redshift and the upper one of the jerk.

%%%%%%%%%%%%%%%%%%%%%%%%%%%%%%%%%%%%%%%%%%%%%%%%%%%%%%%%%%%%%%
\begin{table}
\begin{center}
\begin{scriptsize}
\begin{tabular}{|c|c|c|c|}
\multicolumn{1}{c}{Prior} & \multicolumn{1}{c}{Likelihood} & \multicolumn{1}{c}{$q_0$} & \multicolumn{1}{c}{$z_t$} 
\\\hline
SnIa & CCH & $-0.62^{+0.13}_{-0.11}$ & $0.71^{+0.24}_{-0.14}$  \\ \hline
SnIa & CCH+BAOs & $-0.60\pm 0.10$  & $0.80^{+0.09}_{-0.12}$  \\\hline
CCH & SnIa & $-0.62^{+0.11}_{-0.10}$ & $0.74^{+0.21}_{-0.17}$  \\ \hline
CCH+BAOs & SnIa & $-0.60^{+0.08}_{-0.06}$ & $0.81^{+0.08}_{-0.09}$  \\\hline
\end{tabular}
\end{scriptsize}
\caption{Values of $q_0$ and $z_t$ obtained from the reconstruction of $q(z)$ using the WFR and the expansion (\ref{eq:ExpqY}). We provide the results for four alternative combinations of the prior and the likelihood distributions, see related comments in the main text.}
\end{center}
\label{tab:finaltab}
\end{table}
%%%%%%%%%%%%%%%%%%%%%%%%%%%%%%%%%%%%%%%%%%%%%%%%%%%%%%%%%%%%%% 

Alternatively, we have also explored what happens if we use the expansions (\ref{eq:ExpqZ}) and (\ref{eq:ExpqY}) of $q(z)$ instead of the expansions of $E(z)$ analyzed up to now. First of all we want to remark that if we use these expansions of $q(z)$ the theoretical expressions of the corresponding Hubble rates that enter the fitting analysis loose their linearity in the coefficients. Thus, we cannot obtain the exact analytical expressions for the posterior best-fit values and covariance matrix of the parameters that are used to build $q(z)$. Nevertheless, we can work with the Fisher approximations of the likelihood and posterior distributions (see e.g. \cite{DEbook,AmendolaNotes}) and apply the methodology developed in the last section in order to carry out the reconstructions. In Fig. 6 we plot the reconstructed Hubble rates and deceleration parameters obtained with the WFR method and the expansions (\ref{eq:ExpqZ}) and (\ref{eq:ExpqY}). The theoretical expressions for the Hubble rate are easy to compute using (\ref{eq:q(z)1}). For example, introducing (\ref{eq:ExpqY}) in (\ref{eq:q(z)1}) and integrating the resulting equation we obtain
\begin{equation}
E_J(z)=(1+z)^{1+\sum\limits_{i=0}^{J}q_i}e^{-\sum\limits_{k=1}^{J}\frac{1}{k}\left(\frac{z}{1+z}\right)^k\sum\limits_{i=k}^{J}q_i}
\end{equation}
for $J>0$, and $E(z)=(1+z)^{1+q_0}$ for $J=0$. Analogous expressions can be derived when we use the expansion (\ref{eq:ExpqZ}). To obtain Fig. 6 we have employed the data on CCH+BAOs to build the prior and the Pantheon+MCT data for the likelihood. The left plot shows, again, that the reconstructed Hubble rates obtained from the two expansions under study, (\ref{eq:ExpqZ}) and (\ref{eq:ExpqY}), are fully consistent. Moreover, the reconstructed $q(z)$'s are compatible at the $\sim 1-2\sigma$ c.l. in all the redshift range and the error bands also have a similar size. In this case the current values of the deceleration parameter read $q_0=-0.43^{+0.04}_{-0.07}$ and $q_0=-0.60^{+0.08}_{-0.06}$ for the parametrization (\ref{eq:ExpqZ}) and (\ref{eq:ExpqY}), respectively.

%%%%%%%%%%%%%%%%%%%%%%%%%%%%%%%%%%%%%%%%%%%%%%%%%%%%%%%%%%%%%%
\begin{center}
\begin{figure}
\includegraphics[scale=0.5]{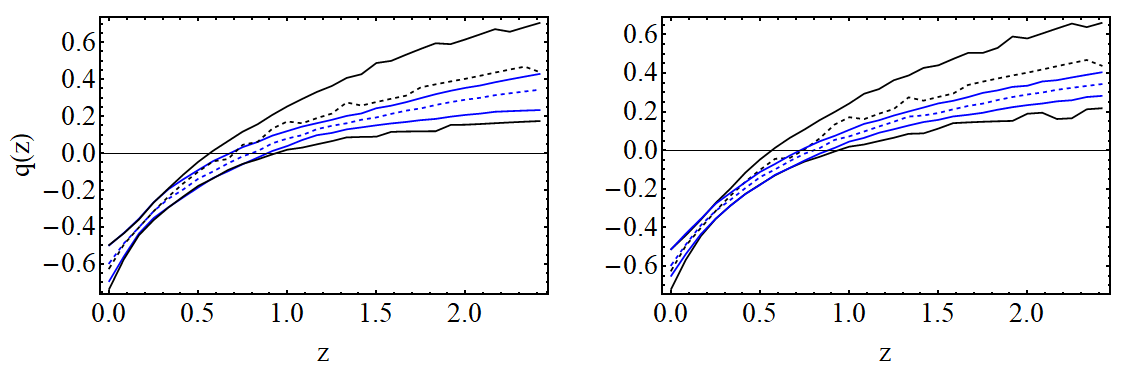}
\caption{{\it Left plot:} Reconstructed deceleration parameters and $1\sigma$-bands obtained with the WFR method and the parametrization (\ref{eq:ExpqY}). The black curves are obtained using the SnIa data in the prior and the CCH in the likelihood. The blue curves by also adding the BAOs data in the likelihood; {\it Right plot:} The same as in the left plot, but using the CCH and CCH+BAOs data in the prior (in black and blue, respectively), and in both cases the SnIa data in the likelihood.}
\end{figure}
\end{center}
%%%%%%%%%%%%%%%%%%%%%%%%%%%%%%%%%%%%%%%%%%%%%%%%%%%%%%%%%%%%%%

We have also studied the differences that are found in the reconstruction of $q(z)$ in the context of the parametrization (\ref{eq:ExpqY}) when one uses the SnIa data in the prior and the CCH/CCH+BAOs in the likelihood instead of using the latter in the prior and the former in the likelihood. The results are shown in Fig. 7. It is evident that this particular choice only has a very minimal impact on our results, the differences are almost imperceptible at naked eye. This can be also checked in Table 6, where we list the values of $q_0$ and the deceleration-acceleration transition redshift $z_t$ that are obtained for the four situations explored in Fig. 7. The latter has been proposed in the literature as a potential primary cosmological parameter, see e.g. \cite{Lima2012}. The constraints are very similar as those obtained with the parametrization (\ref{eq:EexpZ}), and also as those reported in the interesting work \cite{Haridasu2018}. These authors obtained $q_0=-0.52\pm 0.06$ and $z_t=0.64^{+0.12}_{-0.09}$ using the so-called Multi-Task Gaussian Processes technique and a very similar data set as the one employed by us in our analyses, considering the SnIa data of the Pantheon+MCT compilation, CCH and BAOs, as we do. The only differences are the following: they used the BAOs data from \cite{Alam2017,Zhao2019} instead of the data provided in \cite{GilMarin2017,GilMarin2018}; and the CCH data that we have listed in the second column of our Table 2, instead of those listed in the third column of the same table, which also incorporate the systematic errors due to the choice of SPS model and those introduced by the potential presence of a young stellar component in the quiescent galaxies employed in the obtention of the CCH data (cf. Sect. \ref{sec:Data} for details). The reconstructed shape of the deceleration parameter provided in their Fig. 5 is also very similar to those that we have obtained using (\ref{eq:EexpZ}) and (\ref{eq:ExpqY}) in the context of the WFR method, cf. Figs. 4 and 7, and the aforementioned figure in \cite{Haridasu2018}. Thus, there is a very good resonance between the two reconstruction techniques when a reasonable basis for the truncated series is employed in the WFR method\footnote{Notice that despite Gaussian Processes are usually claimed to be non-parametric, one is also forced to choose a reasonable kernel function and to work with the corresponding hyperparameters \cite{GomezAmendola2018}. The choice of such kernel can be thought of as being analogous to the choice of an adequate basis of functions in the WFR method.}. 

We have also reconstructed the jerk parameter $j(z)$ using the expansions (\ref{eq:EexpZ}) and (\ref{eq:ExpqZ}). The results are presented in Fig. 8. They are fully compatible, but unfortunately the errors are still quite large. Nevertheless we can see that the string of data on SnIa+CCH+BAOs employed in this work does not prefer any important deviation from $j(z)=1$, i.e. the predicted value in the $\Lambda$CDM, so there is no need of introducing new physics in order to explain these observations at low and intermediate redshifts. Some authors that have found important evidence in favor of the dynamical nature of the dark energy in the context of various running vacuum and DE models and parametrizations \cite{SGC2017,SCG2018p,SGC2018,SCG2018} have also stated that in order to grasp this dynamical feature one has to use the triad of data on CMB and large-scale structure (especially, those data on redshift space distortions and BAOs that incorporate the matter bispectrum information), and that the current data on SnIa and CCH do not require (when used alone) new physics in the form of dynamical DE. This is in perfect consonance with the results extracted from the reconstructions carried out in this work, and also with the conclusions of \cite{Moews2018}. 
%%%%%%%%%%%%%%%%%%%%%%%%%%%%%%%%%%%%%%%%%%%%%%%%%%%%%%%%%%%%%%

\begin{figure}
\begin{center}
\includegraphics[scale=0.5]{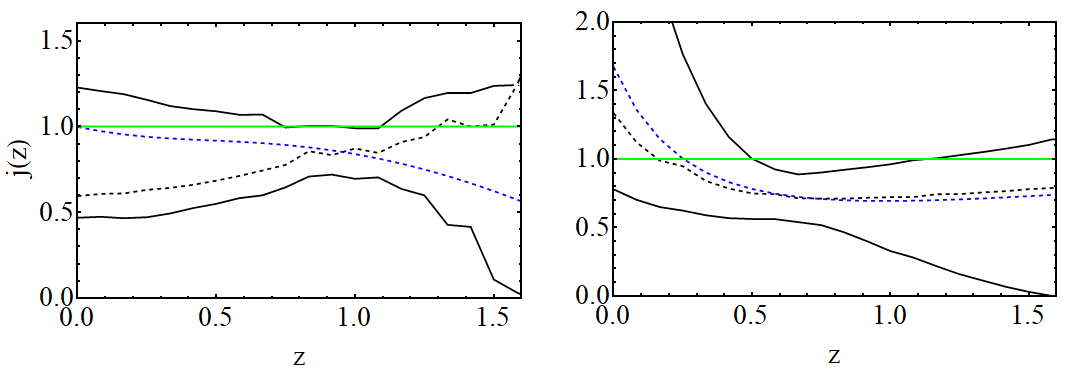}
\caption{{\it Left plot:} Reconstruction of the jerk obtained in the same framework of Fig. 4, using the parametrization for $E(z)$ (\ref{eq:EexpZ}). The error bands cover the $2\sigma$-range. We also plot the $\Lambda$CDM prediction, i.e. $j(z)=1$, in green, and the mean of the reconstructed jerk in blue; {\it Right plot:} The same, but using the parametrization for $q(z)$ (\ref{eq:ExpqZ}). }
\end{center}
\end{figure}

%%%%%%%%%%%%%%%%%%%%%%%%%%%%%%%%%%%%%%%%%%%%%%%%%%%%%%%%%%%%%%

Finally, it is also interesting to compare our more model-independent results, e.g. those compiled in Table 6, with those predicted in the flat $\Lambda$CDM, using the best-fit parameter of $\Omega_m^{(0)}$ obtained from the TT,TE,EE+lowE+lensing+BAO fitting analysis reported by the Planck Collaboration (2018) \cite{Planck2018}, $\Omega_m^{(0)}=0.3111\pm 0.0056$. Using formula (\ref{eq:q0models}) (and setting $w_0=-1$) one obtains $q_0=-0.534\pm 0.008$, and using the $\Lambda$CDM formula for the deceleration-acceleration transition redshift,
\begin{equation}
z_t=\left(\frac{2(1-\Omega_m^{(0)})}{\Omega_m^{(0)}}\right)^{1/3}-1\,,
\end{equation}
one gets $z_t=0.64\pm 0.01$. These values are compatible at $1-2\sigma$ c.l. with those presented in Table 6. Other determinations of $z_t$ obtained using data on $H(z)$ from CCH and BAOs in the context of the $\Lambda$CDM and some dynamical DE models are also consistent with our results, see e.g the works \cite{FarooqRatra,FarooqCrandallRatra,FarooqMadiyar2017}.

%%%%%%%%%%%%%%%%%%%%%%%%%%%%%%%%%%%%%%%%%%%%%%%%%%%%%%%%%%%%%% 
%%%%%%%%%%%%%%%%%%%%%%%%%%%%%%%%%%%%%%%%%%%%%%%%%%%%%%%%%%%%%% 
%%%%%%%%%%%%%%%%%%%%%%%%%%%%%%%%%%%%%%%%%%%%%%%%%%%%%%%%%%%%%%

\section{Conclusions}\label{sec:conclusions}

We have reconstructed in this paper the deceleration and jerk parameters from a very updated data set on SnIa, CCH and BAOs, using the weighted function regression method, and by only assuming the Cosmological Principle and the flatness of the Universe. We have not taken more assumptions for granted, so our analysis can be considered quite model-independent. We have corrected the CCH data in order to incorporate the effect of some systematic uncertainties that are usually disregarded in many other analyses in the literature. We have shown through several examples in Sect. \ref{sec:Motivation} that if we want to infer more objective constraints on the cosmographical functions we are forced not to base our fitting analyses on concrete cosmological models, specific parametrizations of the cosmographical quantities or individual truncated cosmographical expansions. We have studied the correspondence between the latter two scenarios and particular dark energy models in standard GR. The results that we have obtained from our reconstructions are consistent with the standard cosmological model, but the statistical uncertainties associated to the reconstructed functions are much larger (a factor $\sim 3$) than those that are obtained in the context of the $\Lambda$CDM and other particular frameworks. This is actually something expected, and is the price one has to pay for the model-independence. We have computed the level of evidence in favor of the current speed-up of the Universe following a full (and exact) Bayesian approach, using the tools provided in Sect. \ref{sec:WFR}. We have obtained values of $q_0$ which lie $5-6\sigma$ away from $0$ when only the SnIa of the Pantheon+MCT compilation and the CCH are considered. This is in strong contrast with the $17\sigma$ found in the concordance model using the same data. Computing the exact Bayesian evidences and using Jeffreys' scale, we have checked that this corresponds to a moderate level of evidence, whereas it is promoted to a very strong one if also the data on BAOs are taken into account. Thus, these results support the general accepted idea that the Universe is currently undergoing a positive accelerated expansion. This seems to be now something beyond doubt in the light of only low and intermediate-redshift data. Nevertheles, we want to highlight here the importance of carrying out model-independent analyses (when possible) as the one we have reported in this paper to infer unbiased information about the Cosmos, specially when substantial evidence is obtained in the context of concrete cosmological models. We could ask ourselves, for instance, what was the real evidence found in the late nineties by Riess et al. \cite{SNIaRiess} and Perlmutter et al. \cite{SNIaPerl} in favor of an accelerated Universe at the time these seminal papers appeared. Most probably, the $3\sigma$ c.l. evidence reported in these works in the context of the $\Lambda$CDM would have been considerably degraded if a more model-independent method would have been applied to extract the value of $q_0$. This certainly is a more conservative way to proceed. Interestingly, we have also seen that our results are fully compatible with those reported by the authors of \cite{Haridasu2018}, which were obtained using a generalization of the usual Gaussian Processes technique that allowed them to deal with the SnIa+CCH+BAOs data sets in the same analysis simultaneously. In addition, we have provided a new more model-independent determination of the deceleration-acceleration transition redshift, $z_t\sim 0.8\pm 0.1$, and have checked that with the low and intermediate-redshift data sets under consideration the jerk parameter does not require any deviation from the $\Lambda$CDM in order to be explained, although of course, departures from the latter (as e.g. those coming from some sort of dynamical dark energy) are still allowed given the current size of the uncertainties found for this parameter.

It is also very important to remark that although the weighted function regression method improves the standard cosmographical analyses for the reasons mentioned along the paper, it is not completely free from subjective choices. Concretely, one has to select the set of basis functions employed in the reconstruction process. We have explicitly shown that a bad election of this set of basis functions can lead to biased results. This problem is inherent at more or less extent to all the model-independent reconstruction methods, not just to the WFR one. For instance, in order to implement Gaussian Processes one has to choose a concrete kernel function which is in charge of controlling the correlations between different points of the reconstructed function. Regardless of the reconstruction technique one opts to use, one cannot fully escape from these choices. There is no magic solution to this issue, since one has to extrapolate the information carried by the data, at specific points, to construct the continuous function. Hence, the most we can do is trying to minimize the number and impact of our subjective choices. Despite the WFR method still suffers from these subtleties, it must be considered as an important improvement of the usual cosmographical approaches, which allows us to obtain more fair constraints on the cosmological kinematic quantities. In addition, the developed WFR formalism also lets us to compute the Bayes ratio \eqref{eq:BR}, which is a quite objective determination of the statistical (Bayesian) evidence in favor of the current positive acceleration of the Universe.

Nowadays, in the era of precision cosmology we are living, we know that {\it not} all the observational cosmology consists in the search of two numbers, $H_0$ and $q_0$, although these parameters certainly are the ones that can be measured in a more model-independent way with better accuracy, as we have shown in this work. Here we have focused our attention on the second. The measured current value of the deceleration parameter clearly tells us that the Universe is speeding up. The causes of this positive acceleration are still unknown. We think that, in the future, cosmographical analyses as the one carried out in this paper might play an important role to decipher the mystery of the physics behind the current accelerated phase. Despite the jerk parameter is still quite unconstrained by the low and intermediate-redshift data analyzed here, we certainly hope to be capable of putting stringer limits on its value in the coming years, when we have access to more and better data thanks to e.g. the Dark Energy Survey (DES), the Euclid satellite or the Dark Energy Spectroscopic Instrument (DESI). This research line could provide us of new hints about the mechanism that is triggering the current speed-up of the Universe, allowing also the jerk parameter to be a good discriminator of cosmological models, and helping in this way to move towards the solution of one of the most profound enigmas in Physics.

%%%%%%%%%%%%%%%%%%%%%%%%%%%%%%%%%%%%%%%%%%%%%%%
%%%%%%%%%%%%%%%%%%%%%%%%%%%%%%%%%%%%%%%%%%%%%%%

\acknowledgments The author wants to express his gratitude to the Institute of Theoretical Physics of the Ruprecht-Karls University of Heidelberg for the financial support and hospitality during his second short postdoctoral stay there, when the idea of this work came up, and especially to Prof. Luca Amendola for his invitation and the inspiring discussions on the Bayesian evidence held during that time. He is also grateful to Prof. Joan Sol\`a Peracaula for reading this manuscript and for his useful comments on it. To conclude, the author would also like to thank the anonymous Referee for his/her interesting questions and suggestions, which have certainly helped to improve this work.

%%%%%%%%%%%%%%%
%%%%%%%%%%%%%%%


\begin{thebibliography}{200}

\bibitem{SNIaRiess}
A.G. Riess {\it et al.} (Supernova Search Team Collab.), {\it Observational Evidence from Supernovae for an Accelerating Universe and a Cosmological Constant}, Astron. J. {\bf 116} (1998) 1009 [arXiv:astro-ph/9805201]

\bibitem{SNIaPerl}
S. Perlmutter {\it et al.} (Supernova Cosmology Project Collab.), {\it Measurements of $\Omega$ and $\Lambda$ from $42$ High-Redshift Supernovae}, Astrophys. J. {\bf 517} (1999) 565 [arXiv:astro-ph/9812133]

\bibitem{Riess2001} 
A.G. Riess {\it et al.}, {\it The farthest known supernova: support for an accelerating universe and a glimpse of the epoch of deceleration}, Astrophys. J. {\bf 560} (2001) 49 [arXiv:astro-ph/0104455]

\bibitem{TurnerRiess2002} 
M.S. Turner and A.G. Riess, {\it Do SNe Ia provide direct evidence for past deceleration of the universe?}, Astrophys. J. {\bf 569} (2002) 18 [arXiv:astro-ph/0106051]

\bibitem{Knop2003} 
R.A. Knop {\it et al.}, {\it New constraints on Omega(M), Omega(lambda), and w from an independent set of eleven high-redshift supernovae observed with HST }, Astrophys. J. {\bf 598} (2003) 102 [arXiv:astro-ph/0309368] 

\bibitem{Riess2004} 
A.G. Riess {\it et al.}, {\it Type Ia supernova discoveries at $z>1$ from the Hubble Space Telescope: Evidence for past deceleration and constraints on dark energy evolution}, Astrophys. J. {\bf 607} (2004) 665 [arXiv:astro-ph/0402512] 

\bibitem{Sandage1970} 
A. Sandage, {\it Cosmology: a search for two numbers}, Physics Today {\bf 23} (1970) 34.

\bibitem{Weinberg1972} 
S. Weinberg, {\it Gravitation and Cosmology: Principles and Applications of the General Theory of Relativity}, Wiley, New York (1972).

\bibitem{Visser2004} 
M. Visser, {\it Jerk and the cosmological equation of state}, Class. Quant. Grav. {\bf 21} (2004) 2603 [arXiv:gr-qc/0309109] 

\bibitem{Visser2005} 
M. Visser, {\it Cosmography: Cosmology without the Einstein equations}, Gen. Rel. Grav. {\bf 37} (2005) 1541 [arXiv:gr-qc/0411131] 

\bibitem{DunsbyLuongo2016} 
P.K.S. Dunsby and O. Luongo, {\it On the theory and applications of modern cosmography}, Int. J. Geom. Meth. Mod. Phys. {\bf 13} (2016) 1630002 [arXiv:1511.06532]

\bibitem{Sahni2003} 
V. Sahni, T.D. Saini, A.A. Starobinsky and U. Alam, {\it Statefinder: A New geometrical diagnostic of dark energy}, JETP Lett. {\bf 77} (2003) 201; Pisma Zh. Eksp. Teor. Fiz. {\bf 77} (2003) 249 [arXiv:astro-ph/0201498]

\bibitem{Blandford2005} 
R.D. Blandford, M.A. Amin, E.A. Baltz, K. Mandel and P.J. Marshall, {\it Cosmokinetics}, ASP Conf. Ser. {\bf 339} (2005) 27 [arXiv:astro-ph/0408279]

\bibitem{Capozziello2008}
S. Capozziello, V.F. Cardone and V. Salzano, {\it Cosmography of $f(R)$ gravity}, Phys. Rev. D{\bf 78} (2008) 063504 [arXiv:0802.1583]

\bibitem{Tedesco2018} 
L. Tedesco, {\it Ellipsoidal Expansion of the Universe, Cosmic Shear, Acceleration and Jerk Parameter}, Eur. Phys. J. Plus {\bf 133} (2018) 188 [arXiv:1804.11203]

\bibitem{ElgaroyMultamaki2006} 
{\O}. Elgar{\o}y and T. Multam\"{a}ki, {\it Bayesian analysis of friedmannless cosmologies}, J. Cosmol. Astropart. Phys. {\bf 0609} (2006) 002 [arXiv:astro-ph/0603053]

\bibitem{Capozziello2011} 
S. Capozziello, R. Lazkoz and V. Salzano, {\it Comprehensive cosmographic analysis by Markov Chain Method}, Phys. Rev. D{\bf 84} (2011) 124061 [arXiv:1104.3096]

\bibitem{DEbook}
L. Amendola and S. Tsujikawa, {\it Dark energy. Theory and observations}, Cambridge Univ. Press, Cambridge (2010). 

\bibitem{AmendolaNotes} 
L. Amendola, {\it Lecture notes on Statistical Methods}, University of Heidelberg, 2018.

\bibitem{Akaike}
H. Akaike, {\it A new look at the statistical model identification}, IEEE Trans. Autom. Control. {\bf19} (1974) 716.

\bibitem{Schwarz}
G. Schwarz, {\it Estimating the dimension of a model}, Ann. Statist. {\bf 6} (1978) 461.

\bibitem{GomezAmendola2018} 
A. G\'omez-Valent and L. Amendola, {\it $H_0$ from cosmic chronometers and Type Ia supernovae, with Gaussian Processes and the novel Weighted Polynomial Regression method
}, J. Cosmol. Astropart. Phys. {\bf 1804} (2018) 051 [arXiv:1802.01505]

\bibitem{ShapiroTurner2006} 
C. Shapiro and S. Turner, {\it What do we really know about cosmic acceleration?}, Astrophys. J. {\bf 649} (2006) 563 [arXiv:astro-ph/0512586] 

\bibitem{Rapetti2007} 
D. Rapetti, S.W. Allen, M.A. Amin and R.D. Blandford, {\it A kinematical approach to dark energy studies}, Mon. Not. Roy. Astron. Soc. {\bf 375} (2007) 1510 [arXiv:astro-ph/0605683]

\bibitem{GongWang2007} 
Y-G. Gong and A. Wang, {\it Reconstruction of the deceleration parameter and the equation of state of dark energy}, Phys. Rev. D{\bf 75} (2007) 043520 [arXiv:astro-ph/0612196]

\bibitem{Ishida2008} 
E.E.O. Ishida, R.R.R. Reis, A.V. Toribio and I. Waga, {\it When did cosmic acceleration start? How fast was the transition?}, Astrophys. J. {\bf 28} (2008) 547 [arXiv:0706.0546]

\bibitem{CunhaLima2008} 
J.V. Cunha and J.A.S. Lima, {\it Transition Redshift: New Kinematic Constraints from Supernovae}, Mon. Not. Roy. Astron. Soc. {\bf 390} (2008) 210 [arXiv:0805.1261]

\bibitem{Guimaraes2009} 
A.C.C. Guimar\~aes, J.V. Cunha and J.A.S. Lima, {\it Bayesian Analysis and Constraints on Kinematic Models from Union SNIa}, J. Cosmol. Astropart. Phys. {\bf 0910} (2009) 010 [arXiv:0904.3550]

\bibitem{MortsellClarkson2009} 
E. M\"{o}rtsell and C. Clarkson, {\it Model independent constraints on the cosmological expansion rate}, J. Cosmol. Astropart. Phys. {\bf 0901} (2009) 044 [arXiv:0811.0981]

\bibitem{Xu2009} 
L. Xu, W. Li and J. Lu, {\it Constraints on Kinematic Model from Recent Cosmic Observations: SN Ia, BAO and Observational Hubble Data}, J. Cosmol. Astropart. Phys. {\bf 0907} (2009) 031 [arXiv:0905.4552] 

\bibitem{Cunha2009} 
J.V. Cunha, {\it Kinematic Constraints to the Transition Redshift from SNe Ia Union Data}, Phys. Rev. D{\bf 79} (2009) 047301 [arXiv:0811.2379]

\bibitem{Lu2011} 
J. Lu, L. Xu and M. Liu, {\it Constraints on kinematic models from the latest observational data}, Phys. Lett. B{\bf 699} (2011) 246 [arXiv:1105.1871]

\bibitem{GuimaraesLima2011} 
A.C.C. Guimar\~aes and J.A.S. Lima, {\it Could the cosmic acceleration be transient? A cosmographic evaluation}, Class. Quant. Grav. {\bf 28} (2011) 125026 [arXiv:1005.2986] 

\bibitem{Giostri2012} 
R. Giostri {\it et al.}, {\it From cosmic deceleration to acceleration: new constraints from SN Ia and BAO/CMB}, J. Cosmol. Astropart. Phys. {\bf 1203} (2012) 027 [arXiv:1203.3213]

\bibitem{Nair2012}
R. Nair, S. Jhingan and D. Jain, {\it Cosmokinetics: A joint analysis of Standard Candles, Rulers and Cosmic Clocks}, J. Cosmol. Astropart. Phys. {\bf 1201} (2012) 018 [arXiv:1109.4574]

\bibitem{Zhai2013} 
Z-X. Zhai, M-J. Zhang, Z-S. Zhang, X-M. Liu and T-J. Zhang, {\it Reconstruction and constraining of the jerk parameter from OHD and SNe Ia observations}, Phys. Lett. B{\bf 727} (2013) 8 [arXiv:1303.1620]

\bibitem{Akarsu2014} 
\"{O}. Akarsu, T. Dereli, S. Kumar and L. Xu, {\it Probing kinematics and fate of the Universe with linearly time-varying deceleration parameter}, Eur. Phys. J. Plus {\bf 129} (2014) 22 [arXiv:1305.5190] 

\bibitem{Mukherjee2016} 
A. Mukherjee and N. Banerjee, {\it Parametric reconstruction of the cosmological jerk from diverse observational data sets }, Phys. Rev. D{\bf 93} (2016) 043002 [arXiv:1601.05172]

\bibitem{Vargas2016} 
M. Vargas dos Santos, R.R.R. Reis and I. Waga, {\it Constraining the cosmic deceleration-acceleration transition with type Ia supernova, BAO/CMB and $H(z)$ data}, J. Cosmol. Astropart. Phys. {\bf 1602} (2016) 066 [arXiv:1505.03814]

\bibitem{MamonDas2017} 
A.A. Mamon and S. Das, {\it A parametric reconstruction of the deceleration parameter}, Eur. Phys. J. C{\bf 77} (2017) 495 [arXiv:1610.07337]

\bibitem{Jesus2018} 
J.F. Jesus, R.F.L. Holanda, S.H. Pereira, {\it Model independent constraints on transition redshift}, J. Cosmol. Astropart. Phys. {\bf 1805} (2018) 073 [arXiv:1712.01075]

\bibitem{MamonBamba2018} 
A.A. Mamon and K. Bamba, {\it Observational constraints on the jerk parameter with the data of the Hubble parameter}, Eur. Phys. J. C{\bf 78} (2018) 862 [arXiv:1805.02854]

\bibitem{Amirhashchi2018} 
H. Amirhashchi, {\it Recovering $\Lambda$CDM Model From a Cosmographic Study}, [arXiv:1811.05400]

\bibitem{CattoenVisser2007a} 
C. Catto\"{e}n and M. Visser M., {\it Cosmography: Extracting the Hubble series from the supernova data}, [arXiv:gr-qc/0703122]

\bibitem{CattoenVisser2007b} 
C. Catto\"{e}n and M. Visser, {\it The Hubble series: Convergence properties and redshift variables}, Class. Quant. Grav. {\bf 24} (2007) 5985 [arXiv:0710.1887]

\bibitem{Vitagliano2010} 
V. Vitagliano, J-Q. Xia, S. Liberati and M. Viel, {\it High-Redshift Cosmography}, J. Cosmol. Astropart. Phys. {\bf 1003} (2010) 005 [arXiv:0911.1249]

\bibitem{Luongo2011} 
O. Luongo, {\it Cosmography with the Hubble parameter}, Mod. Phys. Lett. A{\bf 26} (2011) 1459

\bibitem{Xu2011} 
L. Xu and Y. Wang, {\it Cosmography: Supernovae Union2, Baryon Acoustic Oscillation, observational Hubble data and Gamma ray bursts}, Phys. Lett. B{\bf 702} (2011) 114 [arXiv:1009.0963]

\bibitem{RubinHayden2016} 
D. Rubin and B. Hayden, {\it Is the expansion of the universe accelerating? All signs point to yes}, Astrophys. J. {\bf 833} (2016) L30 [arXiv:1610.08972]

\bibitem{Dutta2018} 
K. Dutta, Ruchika, A. Roy, A.A. Sen and M.M. Sheikh-Jabbari, {\it Negative Cosmological Constant is Consistent with Cosmological Data}, [arXiv:1808.06623]

\bibitem{Heneka2018} 
C. Heneka, {\it Status of kinematic cosmology with SN Ia: JLA, Pantheon and future constraints with LSST}, [arXiv:1809.04043]

\bibitem{LiDuXu2019}
E-K. Li, M. Du and L. Xu, {\it General Cosmography Model with Spatial Curvature}, [arXiv:1903.11433]

\bibitem{SemizCacimbel2015}
I. Semiz and K. \c{C}amlibel, {\it What do the cosmological supernova data
really tell us?}, J. Cosmol. Astropart. Phys. {\bf 1512} (2015) 038 [arXiv:1505.04043]

\bibitem{Nielsen2016} 
J.T. Nielsen, A. Guffanti and S. Sarkar, {\it Marginal evidence for cosmic acceleration from Type Ia supernovae}, Sci. Rep. {\bf 6} (2016) 35596 [arXiv:1506.01354] 

\bibitem{Ringermacher2016} 
H.I. Ringermacher and L.R. Mead, {\it In Defense of an Accelerating Universe: Model Insensitivity of the Hubble Diagram}, [arXiv:1611.00999]

\bibitem{Haridasu2017} 
B.S. Haridasu, V.V. Lukovi\'c, R. D'Agostino and N. Vittorio, {\it Strong evidence for an accelerating universe}, Astron. Astrophys. {\bf 600} (2017) L1 [arXiv:1702.08244]

\bibitem{Mamon2018} 
A.A. Mamon, {\it Constraints on a generalized deceleration parameter from cosmic chronometers}, Mod. Phys. Lett. A{\bf 33} (2018) 1850056 [arXiv:1702.04916]

\bibitem{Shafieloo2006} 
A. Shafieloo, U. Alam, V. Sahni and A.A. Starobinsky, {\it Smoothing Supernova Data to Reconstruct the Expansion History of the Universe and its Age}, Mon. Not. Roy. Astron. Soc. {\bf 366} (2006) 1081 [arXiv:astro-ph/0505329]
 
\bibitem{Shafieloo2007} 
A. Shafieloo, {\it Model Independent Reconstruction of the Expansion History of the Universe and the Properties of Dark Energy}, Mon. Not. Roy. Astron. Soc. {\bf 380} (2007) 1573 [arXiv:astro-ph/0703034]

\bibitem{Shafieloo2012} 
A. Shafieloo, {\it Crossing Statistic: Reconstructing the Expansion History of the Universe}, J. Cosmol. Astropart. Phys. {\bf 1208} (2012) 002 [arXiv:1204.1109] 

\bibitem{Haridasu2018} 
B.S. Haridasu, V.V. Lukovi\'c, M. Moresco and N. Vittorio, {\it An improved model-independent assessment of the late-time cosmic expansion}, J. Cosmol. Astropart. Phys. {\bf 1810} (2018) 015 [arXiv:1805.03595]

\bibitem{Tutusaus2018} 
I. Tutusaus, B. Lamine and A. Blanchard, {\it Model-independent cosmic acceleration and type Ia supernovae intrinsic luminosity redshift dependence}, Astron. Astrophys. {\bf 625} (2019) A15 [arXiv:1803.06197] 

\bibitem{Riess2017}
A.G. Riess {\it et al.}, {\it Type Ia Supernova distances at redshift $>1.5$ from the Hubble Space Telescope Multy-Cycle Treasury programs: the early expansion rate}, Astrophys. J. {\bf 853} (2018) 126 [arXiv:1710.00844]

\bibitem{Pantheon}
D.M. Scolnic {\it et al.}, {\it The Complete Light-curve Sample of Spectroscopically Confirmed Type Ia Supernovae from Pan-STARRS1 and Cosmological Constraints from The Combined Pantheon Sample}, Astrophys. J. {\bf 859} (2018) 101 [arXiv:1710.00845]

\bibitem{BetouleJLA}
M. Betoule {\it et al.}, {\it Improved cosmological constraints from a joint analysis of the SDSS-II and SNLS supernova samples}, Astron. Astrophys. {\bf 568} (2014) A22 [arXiv:1401.4064]

\bibitem{Metropolis}
N. Metropolis, A. Rosenbluth, M. Rosenbluth, A. Teller and E. Teller, {\it Equation of State Calculations by Fast Computing Machines}, J. Chem. Phys. {\bf 21} (1953) 1087.

\bibitem{Hastings}
W.K. Hastings, {\it Monte Carlo sampling methods using Markov chains and their applications}, Biometrika {\bf57} (1970) 97.

\bibitem{Pinho2018} 
A.M. Pinho, S. Casas and L. Amendola, {\it Model-independent reconstruction of the linear anisotropic stress $\eta$}, J. Cosmol. Astropart. Phys. {\bf 1811} (2018) 027 [arXiv:1805.00027]

\bibitem{Planck2018} 
N. Aghanim {\it et al.} (Planck Collab.), {\it Planck 2018 results. VI. Cosmological parameters}, [arXiv:1807.06209]

\bibitem{Ooba2018} 
J. Ooba, B. Ratra and N. Sugiyama, {\it Planck 2015 Constraints on the Non-flat $\Lambda$CDM Inflation Model}, Astrophys. J. {\bf 864} (2018) 80 [arXiv:1707.03452]

\bibitem{Tutusaus2017}
I. Tutusaus, B. Lamine, A. Dupays and A. Blanchard, {\it Is cosmic acceleration proven by local cosmological probes?}, Astron. Astrophys. {\bf 602} (2017) A73 [arXiv:1706.05036]

\bibitem{JimenezLoeb2002}
R. Jim\'enez and A. Loeb, {\it Constraining Cosmological Parameters Based on Relative Galaxy Ages}, Astrophys. J. {\bf 573} (2002) 37 [arXiv:astro-ph/0106145]

\bibitem{Zhang}
C. Zhang, H. Zhang, S. Yuan, T-J. Zhang and Y-C. Sun, {\it Four new observational H(z) data from luminous red galaxies in the Sloan Digital Sky Survey data release seven}, Res. Astron. Astrophys. {\bf 14} (2014) 1221 [arXiv:1207.4541]

\bibitem{Jimenez}
R. Jim\'enez, L. Verde, T. Treu and D. Stern, {\it Constraints on the equation of state of dark energy and the Hubble constant from stellar ages and the CMB}, Astrophys. J. {\bf 593} (2003) 622 [arXiv:astro-ph/0302560]

\bibitem{Simon}
J. Simon, L. Verde and R. Jim\'enez, {\it Constraints on the redshift dependence of the dark energy potential}, Phys. Rev. D{\bf 71} (2005) 123001 [arXiv:astro-ph/0412269]

\bibitem{Moresco2012}
M. Moresco {\it et al.}, {\it Improved constraints on the expansion rate of the Universe up to $z\sim 1.1$ from the spectroscopic evolution of cosmic chronometers}, J. Cosmol. Astropart. Phys. {\bf 1208} (2012) 006 [arXiv:1201.3609]

\bibitem{Moresco2016}
M. Moresco {\it et al.}, {\it $6\%$ measurement of the Hubble parameter at $z\sim 0.45$: direct evidence of the epoch of cosmic re-acceleration}, J. Cosmol. Astropart. Phys. {\bf 1605} (2016) 014 [arXiv:1601.01701]

\bibitem{Ratsimbazafy2017}
A.L. Ratsimbazafy {\it et al.}, {\it Age–dating Luminous Red Galaxies observed with the Southern African Large Telescope}, Mon. Not. Roy. Astron. Soc. {\bf 467} (2017) 3239 [arXiv:1702.00418]

\bibitem{Stern}
D. Stern, R. Jim\'enez, L. Verde, M. Kamionkowski and S.A. Stanford, {\it Cosmic Chronometers: Constraining the Equation of State of Dark Energy. I: $H(z)$ Measurements}, J. Cosmol. Astropart. Phys. {\bf 1002} (2010) 008 [arXiv:0907.3149]

\bibitem{Moresco2015}
M. Moresco, {\it Raising the bar: new constraints on the Hubble parameter with cosmic chronometers at $z\sim 2$}, Mon. Not. Roy. Astron. Soc. {\bf 450} (2015) L16 [arXiv:1503.01116]

\bibitem{Corredoira2017} 
M. L\'opez-Corredoira, A. Vazdekis, C.M. Guti\'errez and N. Castro-Rodr\'iguez, {\it Stellar content of extremely red quiescent galaxies at $z>2$}, Astron. Astrophys. {\bf 600} (2017) A91 [arXiv:1702.00380]

\bibitem{Corredoira2018} 
M. L\'opez-Corredoira and A. Vazdekis, {\it Impact of young stellar components on quiescent galaxies: deconstructing cosmic chronometers}, Astron. Astrophys. {\bf 614} (2018) A127 [arXiv:1802.09473]

\bibitem{Moresco2018} 
M. Moresco, {\it et al.}, {\it Setting the Stage for Cosmic Chronometers. I. Assessing the Impact of Young Stellar Populations on Hubble Parameter}, Astrophys. J. {\bf 868} (2018) 84 [arXiv:1804.05864] 

\bibitem{BC03}
G. Bruzual and S. Charlot, {\it Stellar population synthesis at the resolution of 2003}, Mont. Not. Roy. Astron. Soc. {\bf 344} (2003) 1000 [arXiv:astro-ph/0309134]
 
\bibitem{MaStro}
C. Maraston and G. Str\"{o}mb\"{a}ck, {\it Stellar population models at high spectral resolution}, Mont. Not. Roy. Astron. Soc. {\bf 418} (2011) 2785 [arXiv:1109.0543]

\bibitem{GilMarin2017}
H. Gil-Mar\'in {\it et al.}, {\it The clustering of galaxies in the SDSS-III Baryon Oscillation Spectroscopic Survey: RSD measurement from the power spectrum and bispectrum of the DR12 BOSS galaxies}, Mon. Not. Roy. Astron. Soc. {\bf 465} (2017) 1757 [arXiv:1606.00439]

\bibitem{Bourboux2017}
H. Mas des Bourboux {\it et al.}, {\it Baryon acoustic oscillations from the complete SDSS-III Ly$\alpha$-quasar cross-correlation function at $z=2.4$},  Astron. Astrophys. {\bf 608} (2017) A130 [arXiv:1708.02225] 

\bibitem{GilMarin2018} 
H. Gil-Mar\'in {\it et al.}, {\it The clustering of the SDSS-IV extended Baryon Oscillation Spectroscopic Survey DR14 quasar sample: structure growth rate measurement from the anisotropic quasar power spectrum in the redshift range 
$0.8<z<2.2$}, Mon. Not. Roy. Astron. Soc. {\bf 477} (2018) 1604 [arXiv:1801.02689]

\bibitem{Planck2015} 
P.A.R. Ade {\it et al.} (Planck Collab.), {\it Planck 2015 results. XIII. Cosmological parameters}, Astron. Astrophys. {\bf 594} (2016) A13 [arXiv:1502.01589]

\bibitem{Verde2017} 
L. Verde, E. Bellini, C. Pigozzo, A.F. Heavens and R. Jimenez, {\it Early Cosmology Constrained}, J. Cosmol. Astropart. Phys. {\bf 1704} (2017) 023 [arXiv:1611.00376]

\bibitem{Heavens2014}
A. Heavens, R. Jim\'enez and L. Verde, {\it Standard rulers, candles, and clocks from the low-redshift Universe}, Phys. Rev. Lett. {\bf 113} (2014) 241302 [arXiv:1409.6217]

\bibitem{RiessH02018}
A.G. Riess {\it et al.}, {\it New Parallaxes of Galactic Cepheids from Spatially Scanning the Hubble Space Telescope: Implications for the Hubble Constant}, Astrophys. J. {\bf 855} (2018) 136 [arXiv:1801.01120]

\bibitem{Cardona2017} 
W. Cardona, M. Kunz and V. Pettorino, {\it Determining $H_0$ with Bayesian hyper-parameters}, J. Cosmol. Astropart. Phys. {\bf 1703} (2017) 056 [arXiv:1611.06088] 

\bibitem{JangLee2017}
S. Jang and M.G. Lee, {\it The Tip of the Red Giant Branch Distances to Type Ia Supernova Host Galaxies. V. NGC 3021, NGC 3370, and NGC 1309 and the Value of the Hubble Constant}, Astrophys. J. {\bf 836} (2017) 74 [arXiv:1702.01118]

\bibitem{Zhang2017}
B.R. Zhang {\it et al.}, \textit{A blinded determination of $H_0$ from low-redshift Type Ia supernovae}, Mon. Not. Roy. Astron. Soc. {\bf 471} (2017) 2254 [arXiv:1706.07573]

\bibitem{Follin2018} 
B. Follin and L. Knox, {\it Insensitivity of the distance ladder Hubble constant determination to Cepheid calibration modelling choices}, Mon. Not. Roy. Astron. Soc. {\bf 477} (2018) 4534 [arXiv:1707.01175]

\bibitem{Riess2011}
A.G. Riess {\it et al.}, {\it A 3\% Solution: Determination of the Hubble Constant with the Hubble Space Telescope and Wide Field Camera 3}, Astrophys. J. {\bf 730} (2011) 119 [{\it Erratum ibid} {\bf 732} (2011) 129] [arXiv:1103.2976]

\bibitem{RiessH02016}
A.G. Riess {\it et al.}, {\it A $2.4\%$ Determination of the Local Value of the Hubble Constant}, Astrophys. J. {\bf 826} (2016) 56 [arXiv:1604.01424]

\bibitem{Addison2018} 
G.E. Addison {\it et al.}, {\it Elucidating $\Lambda$CDM: Impact of Baryon Acoustic Oscillation Measurements on the Hubble Constant Discrepancy}, Astrophys. J. {\bf 853} (2018) 119 [arXiv:1707.06547]

\bibitem{Bonvin2017} 
V. Bonvin {\it et al.}, {\it H0LiCOW – V. New COSMOGRAIL time delays of HE $0435-1223$: $H_0$ to $3.8$ per cent precision from strong lensing in a flat $\Lambda$CDM model}, Mon. Not. Roy. Astron. Soc. {\bf 465} (2017) 4914 [arXiv:1607.01790]

\bibitem{Birrer2018}
S. Birrer {\it et al.}, {\it H0LiCOW - IX. Cosmographic analysis of the doubly imaged quasar SDSS $1206+4332$ and a new measurement of the Hubble constant}, Mon. Not. Roy. Astron. Soc. {\bf 484} (2019) 4726 [arXiv:1809.01274]

\bibitem{Abbott2017} 
B.P. Abbott {\it et al.}, {\it A gravitational-wave standard siren measurement of the Hubble constant }, Nature {\bf 551} (2017) 85 [arXiv:1710.05835]

\bibitem{Guidorzi2017} 
C. Guidorzi {\it et al.}, {\it Improved Constraints on $H_0$ from a Combined Analysis of Gravitational-wave and Electromagnetic Emission from GW170817}, Astrophys. J. {\bf 851} (2017) L36 [arXiv:1710.06426] 

\bibitem{Hotokezaka2018} 
K. Hotokezaka {\it et al.}, {\it A Hubble constant measurement from superluminal motion of the jet in GW170817}, [arXiv:1806.10596]

\bibitem{DadoDar2018} 
S. Dado and A. Dar A., {\it The superluminal motion of the jet launched in GW170817, the Hubble constant, and critical tests of gamma ray bursts theory}, [arXiv:1808.08912]

\bibitem{YuRatraWang2017}
H. Yu, B. Ratra and F-Y. Wang, {\it Hubble Parameter and Baryon Acoustic Oscillation Measurement Constraints on the Hubble Constant, the Deviation from the Spatially-Flat $\Lambda$CDM Model, The Deceleration-Acceleration Transition Redshift, and Spatial Curvature}, Astrophys. J. {\bf 856} (2018) 3 [arXiv:1711.03437]

\bibitem{Feeney2018} 
S.M. Feeney {\it et al.}, {\it Prospects for resolving the Hubble constant tension with standard sirens}, Phys. Rev. Lett. {\bf 122} (2019) 061105 [arXiv:1802.03404]

\bibitem{Lemos2018}
P. Lemos, E. Lee, G. Efstathiou and S. Gratton, {\it Model independent $H(z)$ reconstruction using the cosmic inverse distance ladder}, Mon. Not. Roy. Astron. Soc. {\bf 483} (2018) 4803 [arXiv:1806.06781]

\bibitem{Macaulay2018}
E. Macaulay {\it et al.}, {\it First Cosmological Results using Type Ia Supernovae from the Dark Energy Survey: Measurement of the Hubble Constant}, Mon. Not. Roy. Astron. Soc. {\bf 486} (2019) 2184 [arXiv:1811.02376]

\bibitem{Marra2013} 
V. Marra, L. Amendola, I. Sawicki and W. Valkenburg, {\it Cosmic variance and the measurement of the local Hubble parameter}, Phys. Rev. Lett. {\bf 110} (2013) 241305 [arXiv:1303.3121] 

\bibitem{WuHuterer}
H-Y. Wu and D. Huterer, {\it Sample variance in the local measurements of the Hubble constant}, Mon. Not. Roy. Astron. Soc. {\bf 471} (2017) 4946 [arXiv:1706.09723] 

\bibitem{CamarenaMarra} 
D. Camarena and V. Marra, {\it Impact of the cosmic variance on $H_0$ on cosmological analyses}, Phys. Rev. D{\bf98} (2018) 023537 [arXiv:1805.09900]

\bibitem{Bengaly2018} 
C.A. Bengaly, U. Andrade and J.S. Alcaniz, {\it How does an incomplete sky coverage affect the Hubble Constant variance?}, [arXiv:1810.04966]

\bibitem{Romano2018} 
A.E. Romano, {\it Hubble trouble or Hubble bubble?}, Int. J. Mod. Phys. D{\bf 27} (2018) 1850102 [arXiv:1609.04081]

\bibitem{Keenan2013} 
R.C. Keenan, A.J. Barger and L.L. Cowie, {\it Evidence for a $\sim 300$ Megaparsec Scale Under-density in the Local Galaxy Distribution}, Astrophys. J. {\bf 775} (2013) 62 [arXiv:1304.2884]

\bibitem{Shanks2018a} 
T. Shanks, L.M. Hogarth and N. Metcalfe, {\it GAIA Cepheid parallaxes and ``Local Hole'' relieve $H_0$ tension}, Mon. Not. Roy. Astron. Soc. {\bf 484} (2019) L64 [arXiv:1810.02595]

\bibitem{Shanks2018b} 
T. Shanks, L.M. Hogarth and N. Metcalfe, {\it $H_0$ Tension: Response to Riess et al arXiv:1810.03526}, [arXiv:1810.07628]

\bibitem{RiessProb} 
A.G. Riess, S. Casertano, D. Kenworthy, D. Scolnic and L. Macri, {\it Seven Problems with the Claims Related to the Hubble Tension in arXiv:1810.02595}, [arXiv:1810.03526]

\bibitem{XCDM}
S.M. Turner and M. White, {\it CDM Models with a Smooth Component}, Phys. Rev. D{\bf 56} (1997) R4439 [arXiv:astro-ph/9701138]

\bibitem{CPL} 
M. Chevallier and D. Polarski, {\it Accelerating universes with scaling dark matter}, Int. J. Mod. Phys. D{\bf10} (2001) 213 [arXiv:gr-qc/0009008] 

\bibitem{Linder2003} 
E.V. Linder, {\it Exploring the expansion history of the universe}, Phys. Rev. Lett. {\bf 90} (2003) 091301 [arXiv:astro-ph/0208512]

\bibitem{Linder2004} 
E.V. Linder, {\it Probing gravitation, dark energy, and acceleration}, Phys. Rev. D {\bf 70} (2004) 023511 [arXiv:astro-ph/0402503]

\bibitem{Nesseris2013} 
S. Nesseris and J. Garc\'ia-Bellido, {\it Is the Jeffreys' scale a reliable tool for Bayesian model comparison in cosmology?}, J. Cosmol. Astropart. Phys. {\bf 1308} (2013) 036 [arXiv:1210.7652]

\bibitem{Mathematica}
Wolfram Research, Inc., Mathematica, Version 9.0, Champaign, IL (2012).

\bibitem{KassRaftery1995} 
R.E. Kass and A.E. Raftery, {\it Bayes Factors}, Journal of the American Statistical Association {\bf 90} (1995) 773.

\bibitem{Jeffreys} 
H. Jeffreys, {\it Theory of Probability}, Clarendon Press, Oxford (1961).

\bibitem{Lima2012} 
J.A.S. Lima, J.F. Jesus, R.C. Santos and M.S.S. Gill, {\it Is the transition redshift a new cosmological number?}, [arXiv:1205.4688]

\bibitem{Alam2017} 
S. Alam {\it et al.}, {\it The clustering of galaxies in the completed SDSS-III Baryon Oscillation Spectroscopic Survey: cosmological analysis of the DR12 galaxy sample}, Mont. Not. Roy. Astron. Soc. {\bf 470} (2017) 2617 [arXiv:1607.03155]

\bibitem{Zhao2019} 
G-B. Zhao {\it et al.}, {\it The clustering of the SDSS-IV extended Baryon Oscillation Spectroscopic Survey DR14 quasar sample: a tomographic measurement of cosmic structure growth and expansion rate based on optimal redshift weights}, Mon. Not. Roy. Astron. Soc. {\bf 482} (2019) 3497 [arXiv:1801.03043]

\bibitem{SGC2017} 
J. Sol\`a, A. G\'omez-Valent and J. de Cruz P\'erez, {\it First Evidence of Running Cosmic Vacuum: Challenging the Concordance Model}, Astrophys. J. {\bf 836} (2017) 43 [arXiv:1602.02103]

\bibitem{SCG2018p}
J. Sol\`a, J. de Cruz P\'erez and A. G\'omez-Valent, {\it Dynamical dark energy vs. $\Lambda$=const. in light of observations}, Eur. Phys. Lett. {\bf 121} (2018) 39001 [arXiv:1606.00450]

\bibitem{SCG2018} 
J. Sol\`a, J. de Cruz P\'erez and A. G\'omez-Valent, {\it Possible signals of vacuum dynamics in the Universe}, Mon. Not. Roy. Astron. Soc. {\bf 478} (2018) 4357 [arXiv:1703.08218]

\bibitem{SGC2018} 
J. Sol\`a, A. G\'omez-Valent and J. de Cruz P\'erez, {\it Signs of Dynamical Dark Energy in Current Observations}, Phys. Dark Univ. {\bf 25} (2019) 100311 [arXiv:1811.03505]

\bibitem{Moews2018} 
B. Moews {\it et al.}, {\it Stress testing the dark energy equation of state imprint on supernova data}, [arXiv:1812.09786]

\bibitem{FarooqRatra} 
O. Farooq and B. Ratra, {\it Hubble parameter measurement constraints on the cosmological deceleration-acceleration transition redshift}, Astrophys. J. {\bf 766} (2013) L7 [arXiv:1301.5243]

\bibitem{FarooqCrandallRatra} 
O. Farooq, S. Crandall and B. Ratra, {\it Binned Hubble parameter measurements and the cosmological deceleration-acceleration transition}, Phys. Lett. B{\bf 726} (2013) 72 [arXiv:1305.1957]

\bibitem{FarooqMadiyar2017} 
O. Farooq, F.R. Madiyar, S. Crandall and B. Ratra, {\it Hubble Parameter Measurement Constraints on the Redshift of the Deceleration–acceleration Transition, Dynamical Dark Energy, and Space Curvature}, Astrophys. J. {\bf 835} (2017) 26 [arXiv:1607.03537]



\end{thebibliography}
\end{document}